%% file: cat-decomp.tex
\documentclass[a4paper,UKenglish]{lipics-v2021}

\input{preamble.tex}

\title{Classical simulation of quantum circuits with partial and graphical stabiliser decompositions}
\titlerunning{Classical simulation with partial and graphical stabiliser decompositions}

\author{Aleks Kissinger}{University of Oxford, United Kingdom \and \url{https://www.cs.ox.ac.uk/people/aleks.kissinger/}}{aleks.kissinger@cs.ox.ac.uk}{0000-0002-6090-9684}{Acknowledges support from AFOSR grant FA2386-18-1-4028.}
\author{John van de Wetering}{Radboud University Nijmegen, The Netherlands \and University of Oxford, United Kingdom \and \url{https://vdwetering.name}}{john@vdwetering.name}{https://orcid.org/0000-0002-5405-8959}{NWO Rubicon Personal Grant}
\author{Renaud Vilmart}{Université Paris-Saclay, CNRS, ENS Paris-Saclay, Inria, Laboratoire Méthodes Formelles,\\ 91190, Gif-sur-Yvette, France \and \url{https://rvilmart.github.io}}{vilmart@lsv.fr}{0000-0002-8828-4671}{}

\authorrunning{A.~Kissinger, J.~van de Wetering, R.~Vilmart}

\Copyright{A.~Kissinger, J.~van de Wetering, R.~Vilmart}

\nolinenumbers %uncomment to disable line numbering

\hideLIPIcs  %uncomment to remove references to LIPIcs series (logo, DOI, ...), e.g. when preparing a pre-final version to be uploaded to arXiv or another public repository



\acknowledgements{JvdW and AK would like to thank James Hefford for useful discussions regarding decompositions of cat states in the ZX-calculus.}

\begin{document}

\maketitle

\begin{abstract}
Recent developments in classical simulation of quantum circuits make use of clever decompositions of chunks of magic states into sums of efficiently simulable stabiliser states. 
We show here how, by considering certain non-stabiliser entangled states which have more favourable decompositions, we can speed up these simulations. This is made possible by using the ZX-calculus, which allows us to easily find instances of these entangled states in the simplified diagram representing the quantum circuit to be simulated. 
We additionally find a new technique of \emph{partial} stabiliser decompositions that allow us to trade magic states for stabiliser terms. With this technique we require only $2^{\alpha t}$ stabiliser terms, where $\alpha\approx 0.396$, to simulate a circuit with T-count $t$. This matches the~$\alpha$ found by Qassim et al.~\cite{qassim2021improved}, but whereas they only get this scaling in the asymptotic limit, ours applies for a circuit of any size.
Our method builds upon a recently proposed scheme for simulation combining stabiliser decompositions and optimisation strategies implemented in the software QuiZX~\cite{kissinger2021simulating}.
With our techniques we manage to reliably simulate 50-qubit 1400 T-count hidden shift circuits in a couple of minutes on a consumer laptop.
\end{abstract}

%\section{Introduction and Background}
%
%This paper accompanies Ref.~\cite{kissinger2021simulating} which is attached above.
%Please see the ``Introduction'' and ``Background'' sections of that article for a presentation of the ZX-Calculus, the optimisation strategy used in QuiZX, and the background on stabiliser rank.
%
%QuiZX is an open-source software tool that implements the ZX-diagram optimisation strategy of Ref.~\cite{kissinger2019tcount} used for quantum circuit optimisation. The paper this note relies upon, Ref.~\cite{kissinger2021simulating}, combines this optimisation strategy together with the $6$-magic-states-into-7-stabiliser-terms decomposition of Bravyi, Smith and Smolin~\cite{bravyi2016trading}. We shall refer to this as the \emph{BSS decomposition} in this paper.
%
%We will use the same overall strategy to simulate quantum circuits, but using different stabiliser decompositions. In particular, we still use the optimisation strategy of QuiZX in between every decomposition. This means the diagrams we work with are \emph{reduced}. We recall the following properties of reduced diagrams~\cite{kissinger2021simulating}:
%
%\begin{lemma}
%\label{lem:reduc-clifford-neighbours}
%In a reduced diagram, no two Clifford spiders are neighbours.
%\end{lemma}
%\begin{lemma}
%\label{lem:reduc-binary-clifford}
%In a reduced diagram, no internal Clifford spider can have arity $\leq 2$.
%\end{lemma}
%\begin{lemma}
%\label{lem:reduc-S-spider}
%In a reduced diagram, there are no internal spiders with phase an odd multiple of $\frac\pi2$.
%\end{lemma}

\section{Introduction}

A landmark result in the study of quantum circuit simulation is the Gottesman-Knill theorem~\cite{aaronsongottesman2004}, which states that a quantum circuit consisting of Clifford gates and stabiliser state inputs can be efficiently classically simulated. While Clifford gates are not universal for quantum computation, the Clifford+$T$ gate set, where we also allow the single qubit $T$ gate, can approximate any qubit unitary to arbitrary precision~\cite{aharonov2003simple,ShiToffoliHadamard}. It is widely believed that the Clifford+T gate set requires exponentially large classical resources to simulate in general. However, in practice certain methods exist for simulating Clifford+T circuits significantly larger than one could ever hope to simulate directly, e.g. by state-vector calculation.

% By determining where exactly the frontier between classically simulable resources and speedup-enabling ones lies can help us in providing algorithms for classical simulation of quantum processes. This at the same time allows us to gauge the relative power of quantum computers over classical ones.

A particularly effective technique for circuits with relatively low numbers of non-Clifford gates is simulation by stabiliser decomposition. That is, general states (typically prepared via a Clifford+T circuit) are first decomposed into weighted sums of stabiliser terms, each of which can be simulated efficiently via Gottesman-Knill. The \emph{stabiliser rank} of a state $\ket{\psi}$ is the number of terms in the smallest possible decomposition of $\ket{\psi}$ as a weighted sum of stabiliser states. While computing the exact stabiliser rank of a given state is expected to be a hard problem, various heuristics exist for obtaining stabiliser decompositions which scale with the number of non-Clifford gates in a circuit. By using magic state injection, we can write a state $|\psi\>$ obtained from a Clifford+T circuit as a Clifford circuit that takes as input a $T$ magic state $\ket T := \frac{\ket0+e^{i\frac\pi4}\ket1}{\sqrt2}$ for each $T$ gate in the original circuit. One can then decompose these $T$ magic states, and hence $|\psi\>$ into a sum of stabiliser states. Since each of the individual terms can be simulated efficiently, the simulation cost of the state scales with the number of terms in the decomposition.

Na\"ively, one can decompose each of the $T$ magic states individually as a linear combination of the stabiliser states $|0\>$ and $|1\>$, obtaining $2^t$ terms. However, by decomposing larger non-stabiliser states at once, one can obtain lower simulation costs~\cite{bravyi2016trading,qassim2021improved}. Until now, such techniques have focused on decomposing many identical copies of a fixed magic state.
For example, much work has gone into finding efficient decompositions of $\ket T^{\otimes t}$ for different values of $t$, with the best known decomposition
% In this paper we focus on the exact stabiliser decompositions of states.
scaling asymptoically as $2^{\alpha t}$ with $\alpha\approx 0.396$~\cite{qassim2021improved}.

In previous work by some of the current authors~\cite{kissinger2021simulating}, a new method for simulation with stabiliser decompositions was introduced, which works by representing the circuit to be simulated as a \emph{ZX-diagram}~\cite{CD1,CD2}, and then interleaving the decompositions of~\cite{bravyi2016trading} with the diagram simplification strategy of~\cite{kissinger2019tcount} to reduce the number of stabiliser terms required, sometimes by many orders of magnitude. The scheme goes as follows:
\begin{enumerate}
  \item Translate the circuit, together with the desired input state and measurement effect, to a ZX-diagram.
  \item Simplify the diagram as much as possible using the rules of the ZX-calculus.
  \item \label{item:stabdec} Pick a set of non-Clifford generators and decompose them, obtaining a weighted sum of diagrams with fewer non-Clifford generators.
  \item Apply the previous two steps recursively to each of the diagrams until no non-Clifford generators remain, in which case each diagram is simplified to a single complex number.
  \item The sum of these numbers gives the overall amplitude.
\end{enumerate}
This allowed for the simulation of significantly larger circuits than before. While the previous best simulated a 50-qubit circuit with 64 $T$ gates, in~\cite{kissinger2021simulating} they simulated 50-qubit circuits with up to 1400 $T$ gates.

In this article we build on the work of both~\cite{qassim2021improved} and~\cite{kissinger2021simulating}, by significantly improving Step~\ref{item:stabdec}. 
Our main contribution is to show that it is useful to consider different states than just $\ket T^{\otimes t}$ for decomposition. 
This leads to two improvements. The first is that if certain entangled states are present in the ZX-diagram to be simulated, then we can use a decomposition that asymptotically requires significantly fewer terms, having exponential parameter $\alpha\approx 0.25$ instead of $\alpha\approx 0.396$. While these special states need not appear in a generic ZX-diagram obtained from a Clifford+T circuit, our empirical data suggests that they often will.
It is crucial here that we represent the intermediate stages of the simulation as a ZX-diagram, as opposed to a circuit with magic states as input, as the graph structure makes it possible to find the right entangled states to be decomposed.

Our second improvement is that one of these special decompositions can be adapted to lead to a decomposition of $\ket T^{\otimes 5}$ that uses only 3 terms but leaves one $\ket T$ in the resulting reduced terms, which effectively gives us a `4-to-3' decomposition meaning we require $2^{\alpha t}$ terms with $\alpha\approx 0.396$ in the worst case. This matches the asymptotic upper bound reported in Ref.~\cite{qassim2021improved}, however unlike that paper, we obtain this bound at a fixed finite size rather than in the limit.

We implemented our new decompositions in \texttt{quiZX}, the software used in~\cite{kissinger2021simulating}, and assessed the performance of our method by benchmarking the same quantum circuits as~\cite{kissinger2021simulating}. We found that our new scheme always outperforms the previous one, often by several orders of magnitude. For instance, we benchmarked 125 different 20-qubit hidden-shift circuits with T-count 112. While our method used at most 1 second for each of these, the method of~\cite{kissinger2021simulating} required more than 1000 seconds for some. We also found the distribution of the required time to be a lot less erratic: while 17\% of random hidden shift circuits on 50 qubits with T-count 1400 could be sampled within 5 minutes in the previous proposal, 98\% pass the test with the new decompositions -- and the remaining 2\% only require 1 additional minute, meaning we can reliably simulate 50-qubit hidden-shift circuits with over 1000 T gates on a consumer laptop.

In Section~\ref{sec:zx} we present the ZX-calculus, the formalism used throughout the rest of the paper, and in Section~\ref{sec:stabrank} we discuss the state-of-the-art in stabiliser rank and stabiliser decompositions. In Section~\ref{sec:cat-states} we introduce the decompositions of the entangled states that we will use, as well as a ``partial'' stabiliser decomposition of $\ket T^{\otimes 5}$. Finally, in Section~\ref{sec:benchmarking}, we conduct some benchmarks to assess the relevance of the new decompositions. We end with some concluding remarks in Section~\ref{sec:conclusion}.

\section{ZX-calculus}
\label{sec:zx}

The ZX-calculus is a graphical language for reasoning about quantum computation~\cite{CD1,CD2}. It represents quantum processes using ZX-diagrams which can then graphically be rewritten and simplified using a collection of rewrite rules. For a review see~\cite{vandewetering2020zxcalculus}. Here we give a brief overview of the notions we will need.

ZX-diagrams \cite{CD1} are built from a set of generators: \emph{Z-spiders} represented by green dots, \emph{X-spiders} represented by red dots, the Hadamard gate, represented by a yellow box, and generators capturing ``half-turns'' of a wire and wire crossings. These are defined as follows:
%
% \begin{align*}
% %\small
% \tikzfig{Zsp-a}\quad,\qquad
% \tikzfig{Xsp-a}\quad,\qquad
% \tikzfig{h-alone}
% \end{align*}
%
% Z- and X-spiders can have an arbitrary number of input wires (number of left legs), an arbitrary number of output wires (number of right legs), and an arbitrary phase as a parameter (represented by $\alpha\in\mathbb R$ above). When the phase $\alpha$ is zero we omit it from the spider.
%
% Together with these ZX-specific generators, we have the following ``wire generators'':
% \begin{align*}
% \tikzfig{id}\quad,\qquad
% \tikzfig{swap}\quad,\qquad
% \tikzfig{cap}\quad,\qquad
% \tikzfig{cup}\quad.
% \end{align*}
% ZX-diagrams are interpreted via $\interp{.}$, which maps any ZX-diagram to a complex matrix. This interpretation $\interp{.}$ is inductively defined as follows:

\noindent\begin{minipage}[t]{0.65\columnwidth}
\begin{align*}
\ \input{./figures/Zsp-a.tikz}\  &= \ \ketbra{0...0}{0...0} +
e^{i \alpha} \ketbra{1...1}{1...1} \\
\ \input{./figures/Xsp-a.tikz}\  &= \ \ketbra{+...+}{+...+} +
e^{i \alpha} \ketbra{-...-}{-...-} \\
\input{./figures/h-alone.tikz} &= \ \ketbra+0 +\ketbra-1
\end{align*}
\end{minipage}
\hfill
\begin{minipage}[t]{0.34\columnwidth}
\begin{align*}
% \interp{D_2\otimes D_1} &= \ \interp{D_2}\otimes\interp{D_1} \\
\input{./figures/id.tikz} &= \ \ketbra00+\ketbra11\\
\input{./figures/swap.tikz} &= \ \sum_{i,j\in\{0,1\}}\ketbra{ij}{ji}\\
\input{./figures/cap.tikz}~ &= \ \ket{00}+\ket{11}\\
~\input{./figures/cup.tikz} &= \ \bra{00}+\bra{11}
\end{align*}
\end{minipage}

\bigskip

Note that, much like in quantum circuit notation, we interpret composition as ``plugging'' diagrams together and tensor product as putting diagrams side-by-side:
\[
  \left(\input{./figures/D2.tikz}\right)\circ\left(\input{./figures/D1.tikz}\right) = \input{./figures/D1.tikz}\!\input{./figures/D2.tikz} \qquad\quad
  \left(\input{./figures/D1.tikz}\right)\otimes\left(\input{./figures/D2.tikz}\right) = \begin{array}{c}\input{./figures/D1.tikz}\\[0.5em]\input{./figures/D2.tikz}\end{array}
\]

We can hence build complicated diagrams by composition of these small bricks, and interpret them as a (not necessarily unitary) linear map. For example, every quantum circuit built from the traditional gate set $\langle \CNOT,H,Z_\alpha\rangle$ can be directly mapped into a ZX-diagram via the translation:
\begin{align*}\label{eq:zx-gates}
\CNOT & \mapsto \sqrt{2}\ \input{./figures/cnot-aux.tikz} &
Z_\alpha & \mapsto \input{./figures/Z-a.tikz} &
H & \mapsto \input{./figures/h-alone.tikz}
\end{align*}
Notice that we used here a complex number explicitly as a global multiplicative scalar to the CNOT diagram. These scalars can be represented in ZX, but are a lot more convenient to deal with explicitly when possible. Any diagram with 0 inputs and 0 outputs represents a scalar, and when it can be efficiently computed, we will simply do this, instead of leaving it as a diagram. 

An important subclass of ZX-diagrams are the \textit{Clifford ZX-diagrams}, i.e. those whose spiders only have phases equal to integer multiples of $\pi/2$. A Clifford ZX-diagram with no input wires always represents a stabiliser state, and conversely any stabiliser state can always be written as a Clifford ZX-diagram~\cite{BackensCompleteness}.

% These efficiently computed scalars include:
% $$\interp{\tikzfig{sqrt-2}}=\sqrt2~,\quad\interp{\tikzfig{1_sqrt-2}}=\frac1{\sqrt2}~,\quad\interp{\ \tikzfig{phase}\ }=\sqrt2 e^{i\alpha}~,\quad\interp{\ \tikzfig{Z-scalar}\ }=1+ e^{i\alpha}$$

A convenient feature of ZX-diagrams is that the linear map they represent only depends on the connectivity of the diagram and not the actual positions of spiders or the direction of wires between them. This means in particular that we can topologically deform a diagram however we wish without changing its interpretation. This allows us to treat a ZX-diagram as an undirected open (multi-)graph, whose vertices are Z-spiders, X-spiders and H-gates. As H-gates are binary, it can be convenient to internalise them on their respective wires, and consider a new wire type which we call an \emph{H-edge}, represented by a dashed blue line:
\ctikzfig{blue-edge-def} 

In addition to topological deformation, there are graphical rewrite rules that also preserve the semantics of the diagram, and in fact,
\emph{complete} axiomatisations of the ZX-calculus exist (e.g.~\cite{HarnyAmarCompleteness,SimonCompleteness,vilmarteulercompleteness}), which entirely capture the semantical equivalence of diagrams. We will not give a full such axiomatisation here, but only those rules we will need in this paper:
\[\input{./figures/spider-rule.tikz}\quad,\qquad
\begin{array}{c}\input{./figures/h-involution-rule.tikz}\\[1em]\input{./figures/0-rotation.tikz}\end{array}
\quad,\qquad\input{./figures/colour-change-rule.tikz}\]
which by duality remain true when Z- and X-spiders are interchanged (i.e.~when the colours are swapped).

Using these rules, it is possible to turn any diagram in a form where i) all spiders are Z-spiders and ii) all edges are H-edges (but inputs and outputs remain plain wires). A diagram in such a form is called \emph{graph-like}~\cite{cliffsimp}. These are the diagrams we will work with in this paper.

While the rewrite rules are unidirectional, we can apply one-way rewrite strategies that try to reduce some metric of the diagram, such as the number of spiders in a diagram.
In particular, in~\cite{cliffsimp} a simplification strategy was introduced that can remove most Clifford spiders, i.e. those spiders whose phase is an integer multiple of $\pi/2$, from a ZX-diagram. It is this rewrite strategy (or more specifically, the improved version from~\cite{kissinger2019tcount}) that was used in~\cite{kissinger2021simulating} to help in simplifying diagrams for simulation with stabiliser decompositions. As the details are not relevant to us, we will only describe some of the properties that the resulting diagrams enjoy, as these will be important for our results. We will call diagrams simplified by the rewrite strategy of~\cite{kissinger2019tcount} \emph{reduced}. Note that a spider is called \textit{internal} if it is not directly connected to an input or an output of the diagram.

\begin{lemma}
\label{lem:reduc-clifford-neighbours}
In reduced diagrams, no two Clifford spiders are neighbours.
\end{lemma}
In other words, patterns like \input{./figures/neighbouring-cliffords.tikz} won't appear in the diagram.
\begin{lemma}
\label{lem:reduc-binary-clifford}
In reduced diagrams, no internal Clifford spider can have arity $\leq 2$.
\end{lemma}
In other words, patterns like \input{./figures/binary-clifford.tikz} and \input{./figures/unary-clifford.tikz} won't appear.
\begin{lemma}
\label{lem:reduc-S-spider}
In reduced diagrams, no internal spider phases are odd multiples of~$\frac\pi2$.
\end{lemma}
In other words, patterns like \input{./figures/internal-S-spider.tikz} won't appear.

% Two restricted classes of ZX-diagrams (that are in correspondence with their circuits' counterparts) are of importance for the remainder of the article.
% \begin{itemize}
% 	\item The \emph{Clifford fragment} which consists of diagrams where the phases are integer multiples of $\frac\pi2$. By~\cite{aaronsongottesman2004}, these diagrams are efficiently simulable by classical means.
% 	\item The \emph{Clifford+T fragment} where phases are integer multiples of $\frac\pi4$. These diagrams can represent the approximately universal Clifford+T gate set.
% \end{itemize}
% In the rest of the paper we will find ways to write Clifford+T diagrams as weighted sums of Clifford diagrams in order to better classically simulated Clifford+T quantum circuits.

\section{Stabiliser Rank and stabiliser decompositions}
\label{sec:stabrank}

Since stabiliser states/operators are efficiently classically simulable, it is worthwhile to try to find decompositions of arbitrary quantum states as linear combination of stabiliser states with a small number of terms.

More formally, let $\ket\psi$ be an arbitrary state. Then a \emph{stabiliser decomposition} of $\ket\psi$ is a decomposition $\ket\psi=\sum_{k}^n \lambda_k \ket{\psi_k}$ where $\ket{\psi_k}$ are all stabiliser states and $\lambda_k\in \mathbb{C}$ are some scalars. The smallest $n$ for which such a decomposition exists is called the stabiliser rank of $\ket\psi$, and is denoted by $\chi(\ket\psi)$. For simulation purposes, we are obviously not only interested in $\chi(\ket\psi)$ but also in an associated decomposition.

In the Clifford+T case, most of the results on stabiliser decompositions have focussed on tensor products of the magic state $\ket T:=\frac{1}{\sqrt{2}}(\ket0+e^{i\frac\pi4}\ket1)$, which we can represent as a ZX-diagram (up to normalisation) by \input{./figures/magic-state.tikz}. 
Whenever we have a spider with phase an odd multiple of $\pi/4$ we can unfuse a magic state:
\[\input{./figures/T-spider-to-magic-state.tikz}\]

A single magic state can be decomposed as a sum of two stabiliser states:
\[\input{./figures/magic-state.tikz} = \frac1{\sqrt2}\left(\input{./figures/ket-0-GL.tikz}+e^{i\frac\pi4}\input{./figures/ket-1-GL.tikz}\right)\]
A naive decomposition of $\input{./figures/magic-state.tikz}^{\otimes t}$ would then be to apply this decomposition for all magic states, which would result in a decomposition with $2^t$ terms. More sophisticated decompositions exist however, such as:
\[\input{./figures/2-T-decomposition.tikz}\]
We can use this decomposition to reduce $T$ magic states pairwise, bringing the total number of terms down to $2^{\frac t2}$. 
There is also a $\ket T^{\otimes 6}$ decomposition requiring 7 terms, described in~\cite{bravyi2016trading} and used in~\cite{kissinger2021simulating}. More recently, an improved decomposition of $\ket T^{\otimes 6}$ was found that only requires 6 terms~\cite{qassim2021improved}.
To compare different decompositions it will be helpful to consider how many stabiliser terms would be needed if we were to decompose every non-Clifford phase in the diagram using that decomposition. 
If a decomposition reduces the T-count by $r$ using $p$ terms, we would need $p^{t/r} = 2^{\alpha t}$ stabiliser terms. This $\alpha = \log_2(p)/r$ will be the metric we want to minimise.
For instance, the decomposition of~\cite{bravyi2016trading} gives $\alpha\approx0.468$ ($r=6$, $p=7$).

In the following, we will consider some entangled non-stabiliser states of which we can find a better decomposition than would be suggested by the number of magic states needed to write down these states.

\section{Results}
\label{sec:cat-states}

We present two different improvements to the stabiliser decompositions used before: using decompositions of certain entangled states instead of product states, and using `partial' stabiliser decompositions where the terms are not Clifford, but merely contain a lower amount of magic states.

\subsection{Cat states in the ZX-calculus}

%TODO: check scalars

%In previous works using stabiliser decompositions for quantum circuit simulation, the states that are usually considered for decomposition are tensor products of the magic state $\ket T:=\frac{1}{\sqrt{2}}(\ket0+e^{i\frac\pi4}\ket1)$, represented as a ZX-diagram (up to normalisation) by \tikzfig{magic-state}. Such a magic state corresponds in the ZX-calculus to any spider which has an odd multiple of $\pi/4$ as a phase:
%\[\tikzfig{T-spider-to-magic-state}\]
%In the following, we will consider some entangled non-stabiliser states whose stabiliser rank we know a better upper bound of than would be suggested by the number of magic states needed to write down these states.
%
%To compare different decompositions it will be helpful to consider how many stabiliser terms would be needed if we were to decompose every non-Clifford phase in the diagram using that decomposition. 
%If a decomposition reduces the T-count by $r$ using $p$ terms, we would need $p^{t/r} = 2^{\alpha t}$ stabiliser terms. This $\alpha = \log_2(p)/r$ will be the metric we want to minimise.
%For instance, the decomposition of~\cite{bravyi2016trading} gives $\alpha\approx0.468$ ($r=6$, $p=7$).

The states we are interested in here were introduced in Ref.~\cite{qassim2021improved} as \emph{cat states}: 
\[\ket{\operatorname{cat}_n}:= \frac1{\sqrt2} (\mathbb I^{\otimes n} + Z^{\otimes n})\ket T^{\otimes n} = \frac1{\sqrt2^{n+1}}\left(\input{./figures/magic-state.tikz}^{\otimes n}+\input{./figures/magic-state-bot.tikz}^{\otimes n}\right)\]
In Ref.~\cite{qassim2021improved}, these states are used because i) they have a relatively small stabiliser rank and ii) because if the stabiliser rank of $\ket{\operatorname{cat}_n}$ is bounded by $c$, then that of $\ket T^{\otimes n}$ is bounded by $2c$.
% , thanks to the identity $\ket T^{\otimes n} = \frac12\ket{\operatorname{cat}_n}+\frac{e^{i\frac\pi4}}2(Z^{(1,1)}(\frac\pi2)X^{(1,1)}(0)\otimes \mathbb I^{\otimes n-1})\ket{\operatorname{cat}_n} = \frac12\ket{\operatorname{cat}_n}+\frac{e^{i\frac\pi4}}2\left(\tikzfig{SoX}\right)\ket{\operatorname{cat}_n}$. 
For instance, they find a decomposition of $\ket{\operatorname{cat}_6}$ using 3 stabiliser states, which hence gives a decomposition of $\ket T^{\otimes6}$ using 6 stabiliser states. This improves upon the 6-to-7 decomposition of Ref.~\cite{bravyi2016trading}, and improves $\alpha\approx 0.468$ to $\alpha\approx 0.431$.

In this paper we focus on the use of the cat-states directly for stabiliser decompositions, rather than first transforming them into decompositions of sets of $T$ states. First, notice that they can be easily expressed as ZX-diagrams:
\[\ket{\operatorname{cat}_n} = \frac1{\sqrt2}~ \scalebox{0.8}{\input{./figures/cat-state.tikz}}\]
%(or even as circuits as: \tikzfig{cat-state-circuit}).

We can then translate to ZX the decomposition of $\ket{\operatorname{cat}_6}$ found in Ref.~\cite{qassim2021improved}%
\footnote{The description of the last diagram in the decomposition is not the one given in \cite{qassim2021improved}. However, the two can be shown to be equivalent using the ZX-calculus.}%
:
\[\scalebox{0.9}{\input{./figures/cat-6-decomp.tikz}}\]
Hence, if our diagram contained enough subdiagrams that look like $\ket{\operatorname{cat}_6}$ to allow us to decompose all non-Clifford spiders using this decomposition we would get $3^{\frac t6}=2^{\alpha t}$ terms, where $\alpha \approx 0.264$. 
We can in fact do even better by considering the decomposition of $\ket{\operatorname{cat}_4}$, which has a stabiliser rank of only $2$: \def\fig{cat4}
\[\scalebox{0.9}{\input{./figures/cat-4-decomp.tikz}}\]
If we could use only this decomposition we would get $2^{0.25t}$ terms.

The existence of these beneficial decompositions suggests a new strategy: at every step of the decomposition of the complete diagram, we look at every internal spider in the graph and the shape of its neighbourhood and we pick a target for decomposition that requires the fewest number of terms per magic state. After each decomposition, we use a ZX-diagram simplification strategy (like the one in~\cite{kissinger2021simulating}) and then we repeat the procedure.
%so we can use Lemmas~\ref{lem:reduc-clifford-neighbours},~\ref{lem:reduc-binary-clifford} and~\ref{lem:reduc-S-spider}.

We cannot hope to always have an occurrence of $\ket{\operatorname{cat}_4}$ or $\ket{\operatorname{cat_6}}$ at hand. However, notice that we can always infer a decomposition of $\ket{\operatorname{cat}_n}$ from one of $\ket{\operatorname{cat}_{n+k}}$ ($k>0$), by $\ket{\operatorname{cat}_{n}} = (\bra0^{\otimes k}\otimes \mathbb I^{\otimes n})\ket{\operatorname{cat}_{n+k}}$. This observation gives us for instance a decomposition of $\ket{\operatorname{cat}_3}$ and $\ket{\operatorname{cat}_5}$:
\[\scalebox{0.85}{\input{./figures/cat-3-decomp.tikz}} \qquad\quad \scalebox{0.85}{\input{./figures/cat-5-decomp.tikz}}\]
These two decompositions give us respectively $\alpha=1/3\approx 0.333$ and $\alpha \approx 0.317$.

Note that by Lemma~\ref{lem:reduc-binary-clifford}, there will never be any occurrence of $\ket{\operatorname{cat}_2}$ or $\ket{\operatorname{cat}_1}$ in our simplified diagrams so that we now have a strategy for decomposing any $\ket{\operatorname{cat}_n}$ for $n\leq6$.
% Right away, we suggest to give priority to Clifford spiders: we shan't consider T-spiders for decomposition if Clifford spiders are still present. Notice that by Lemma \ref{}, all the neighbouring nodes of a Clifford spider are T-spiders.

%The remaining situations are then of three sorts: 
%\begin{enumerate}
%\item the considered node has more than 6 neighbouring T-spiders
%\item the considered node is itself a T-spider
%\item both of the above
%\end{enumerate}
%
%In cases 1. and 3., we suggest to use the following pre-processing that allows us to keep using cat-state decompositions:
%\[\tikzfig{cat-more-6}\]
%Notice that this preprocessing artificially creates $2$ T-spiders, and that $6$ will be removed by decomposition of $\ket{\operatorname{cat}_6}$. We hence globally remove $4$ T-spiders using 3 terms. This is still advantageous to magic-state decompositions, as $\frac{\log_2(3)}{4}\simeq 0.396 < 0.431\simeq \frac{\log_2(6)}{6}$.
%
%In the case 2. (and with fewer than 6 neighbours), we can do the following preprocessing:
%\[\tikzfig{cat-T}\]
%However, this is only advantageous if the initial number of neighbours is 5. Indeed, in that case we again get an $\alpha$ of $\frac{\log_2(3)}4$, but already for $4$ neighbours, we only remove $3$ T-spiders at the cost of introducing $3$ new terms, which gives $\alpha = \frac{\log_2(3)}3\simeq 0.528$. It can be verified that for fewer neighbours, we still have $\alpha > \frac{\log_2(6)}6$.

These decompositions apply when the ``central'' Clifford spider has phase $0$. By Lemma~\ref{lem:reduc-S-spider} any internal Clifford spider will have phase $0$ or $\pi$. The $\pi$ case is easily taken care of by pushing the phase through one of the neighbours as follows:
\[\input{./figures/pi-cat-state.tikz}\]

Note that in practice, in our reduced diagrams the only occurrences of cat-states will be as \emph{phase gadgets}~\cite{kissinger2019tcount,phaseGadgetSynth}, which relate to cat states in the following way:
\[\input{./figures/phase-gadget-to-cat.tikz}\]
% \[\tikzfig{phase-gadget-in-situ}\]
% We can write these in the following way:
Hence, an $n$-legged phase gadget corresponds to a $\ket{\text{cat}_{n+1}}$ state, so that it can be decomposed more efficiently.

In Ref.~\cite{qassim2021improved} they also describe ways to get efficient stabiliser decompositions of $\ket{\text{cat}_n}$ for bigger $n$. For instance, they find a decomposition of $\ket{\text{cat}_{10}}$ which has $9$ terms, which gives $\alpha\approx 0.317$. For simplicity of implementation, we don't consider these decompositions of bigger states here.

\subsection{Partial magic state decompositions}

If we cannot find cat states in our diagram (which means that it only has non-Clifford spiders or that every Clifford spider has arity larger than 6), then we could fall back to the magic state decompositions described in Ref.~\cite{qassim2021improved}.
Namely, as is observed in Ref.~\cite{qassim2021improved}, we can construct decompositions of bigger cat-states by the following identity:
\[\input{./figures/cat-n.tikz}\]
Keep in mind that $\ket{\operatorname{cat}_{n}} = (\bra0^{\otimes k}\otimes \mathbb I^{\otimes n})\ket{\operatorname{cat}_{n+k}}$ so that we also get decompositions for~$n$ that are not equal to $4k+2$.
Using this construction of $\ket{\operatorname{cat}_{4k+2}}$, we get a decomposition of $\ket T^{\otimes 4k+2}$ that uses $2\cdot 3^k$ terms. This decomposition requires $2^{\alpha (4k+2)}$ terms where
\[\alpha = \frac{1+k\log_2(3)}{4k+2}\underset{k\to\infty}\longrightarrow \frac{\log_2(3)}4\approx 0.396.\]

But as we will see now, we can in fact reach this asymptotic $\alpha$ without needing to decompose all magic states at once. The idea here is to use the decomposition of $\ket{\operatorname{cat}_6}$ to reduce the T-count of the diagram by 4, using only 3 terms. This can be done whenever we have a T-count $\geq5$, in the following way:
\[\scalebox{0.9}{\input{./figures/magic-state-5-decomp.tikz}}\]
%\begin{align*}
%\tikzfigc{00}
%\eq{}4\tikzfigc{01}
%\eq{}2\tikzfigc{02}\\
%+2\sqrt2e^{i\frac34\pi}\tikzfigc{03}
%+512e^{i\frac54\pi}\tikzfigc{05}
%\end{align*}
This is a (non-stabiliser) decomposition of the 5-qubit magic state which leaves one T-spider in each term. So each term effectively loses 4 magic states, so that we can view this as a 4-to-3 strategy. We get for this operation $\alpha = \frac{\log_2(3)}4\approx 0.396$ so that we reach this number concretely that in Ref.~\cite{qassim2021improved} was only reached asymptotically.

Additional advantages of this decomposition over the strategy using the $\ket{\operatorname{cat}_{4k+2}}$ decomposition, are that i) it is simpler to handle and implement (as a smaller part of the diagram is involved) and ii) it more easily allows for the ZX-diagram reduction strategy of~\cite{kissinger2021simulating} to be used in-between rounds of decompositions, so that there is more possibility for reduction of the number of terms.

We remark that this type of `partial decomposition' can be used to decompose any $\ket{T}^{\otimes 4k+1}$ states using the $\ket{\operatorname{cat}_{4k+2}}$ decomposition, but doing so always results in the same $\alpha$, whenever $k\geq1$, so that there is no benefit to doing so (at least from an asymptotic perspective).

\subsection{The full decomposition strategy}

The conclusion of the previous sections is that some subdiagrams are better to decompose into stabilisers than others.
In particular, in our diagrams we should look, in order of importance, for:
\begin{enumerate}
\item a Clifford spider with 4 neighbours ($\alpha = 0.25$),
\item a Clifford spider with 6 neighbours ($\alpha \approx 0.264$),
\item a Clifford spider with 5 neighbours ($\alpha \approx 0.317$),
\item a Clifford spider with 3 neighbours ($\alpha =1/3 \approx 0.333$),
\item any 5 T-spiders ($\alpha \approx 0.396$).
\end{enumerate}

Once the best subdiagram to decompose has been found, we apply the associated decomposition, then simplify each resulting diagram using the optimisation strategy of~\cite{kissinger2019tcount} as described in~\cite{kissinger2021simulating}, and repeat the procedure for each term until we are left with diagrams that have fewer than 5 T-spiders and no cat-states, in which case we simply fall back to the usual magic-state decompositions.

\section{Benchmarking}
\label{sec:benchmarking}

To assess the effectiveness of our approach, we performed simulations of the two families of circuits that were benchmarked in~\cite{kissinger2021simulating}, and compare the associated running times with theirs. For a better comparison, we turned off parallelisation in most experiments. All subsequent tests were performed on a consumer laptop, equipped with an Intel Core i5-10400H CPU, 2.60GHz. By using a depth-first decomposition approach the amount of memory needed for these benchmarks was insignificant. In these benchmarks we refer to to the approach of~\cite{kissinger2021simulating} as \emph{BSS}, which stands for Bravyi, Smith, Smolin, as it uses the decomposition introduced in~\cite{bravyi2016trading}.

\subsection{Random Clifford+T circuits}

The first family of circuits we consider are Clifford+T circuits. We construct these by composing a given number (the T-count $t$) of \emph{exponentiated Pauli unitaries}, of the form $\exp(-i(2{k}+1)\frac\pi4P)$ where $P$ is a Pauli string, i.e. a tensor product of 1-qubit Pauli operators. The \emph{weights} of the Pauli strings (the number of non-identity operators) is chosen randomly between~$2$ and~$4$, in order to mimic the structure common to quantum chemistry circuits, where the Hamiltonian has terms of weight $2$--$4$.

\begin{figure}[!htb]
\centering
\includegraphics[width=1.0\columnwidth]{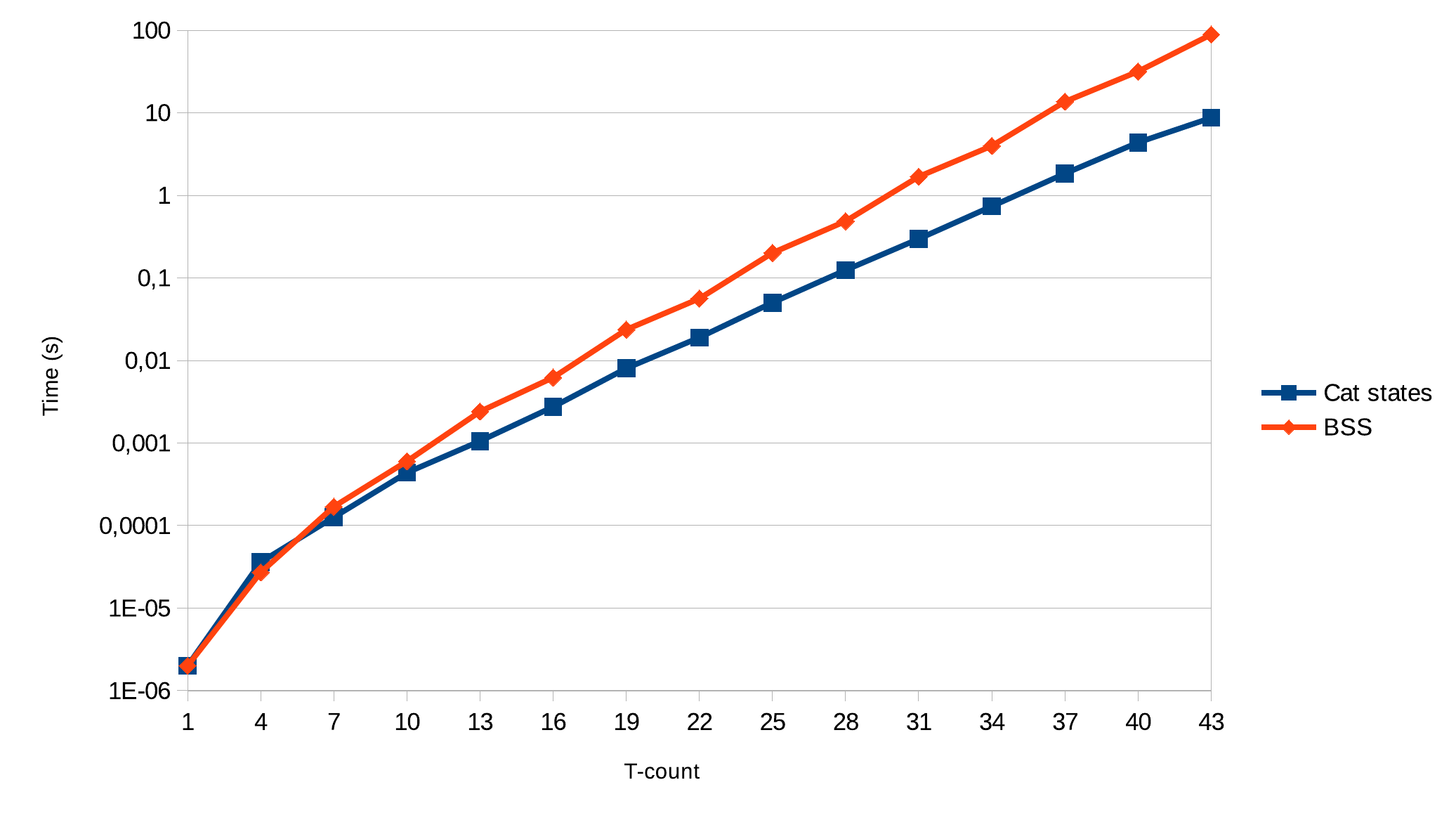}
\vspace{-0.6cm}
\caption{Runtime of random 20-qubit Clifford+T circuit simulations (avg of 10 runs per T-count).}
\label{fig:random-cliffordT}
\end{figure}

We considered random 20-qubit Clifford+T circuits with T-counts varying from $1$ to $43$ (increasing in steps of $3$). We calculated the amplitude of these circuits for a fixed input of $\ket{+}^{\otimes 20}$ and output of $\bra{+}^{\otimes 20}$ (where $\ket+:=\frac1{\sqrt2}(\ket0+\ket1)$) using the two decomposition strategies. 
The running times shown in Figure~\ref{fig:random-cliffordT} are averages over 10 runs for each T-count. We see already an order of magnitude improvement between the running time of the two approaches for a T-count of 43.

\subsection{Random hidden-shift circuits}

The second family of circuits we consider are hidden-shift circuits, whose description can be found in Refs.~\cite{Bravyi2019simulationofquantum,bravyiImprovedClassicalSimulation2016}. These circuits are composed of H, Z, CZ and CCZ gates. The first three are readily translatable to ZX-diagrams, while the CCZ gate requires a decomposition that introduces T-spiders. The decomposition used here is one that uses 7 T-spiders for each CCZ gate.

\begin{figure}[!htb]
\centering
\includegraphics[width=1.0\columnwidth]{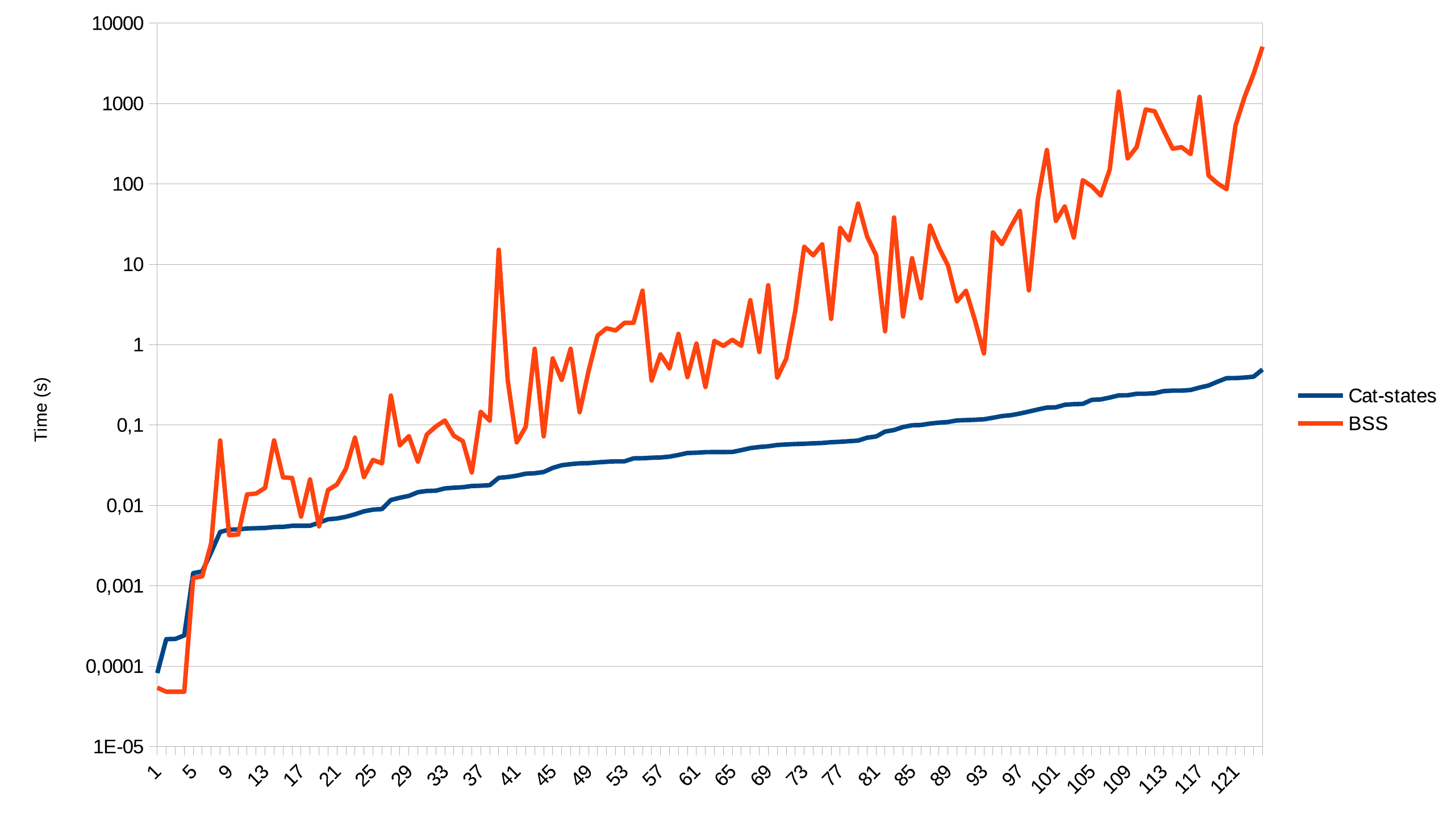}
% \vspace{-0.2cm}
\caption{Logarithmic plot of the runtimes of simulating 125 different 20-qubit hidden-shift circuits with T-count 112, using the BSS decomposition and our strategy. Each datapoint corresponds to a circuit and they are sorted left-to-right according to simulation time using our approach.}
\label{fig:hidden-shift-sorted}
\end{figure}

We found the cat-state decompositions to be particularly efficient compared to the BSS decomposition for this class of circuits. We empirically found the BSS-based approach to be quite erratic, with running times varying between 0.005s and more than 1000s for circuits with the same T-count. This aspect seems to be dampened when using cat-state decompositions. To better gauge this phenomenon, we generated 125 20-qubit hidden-shift circuits with T-count 112, and simulated all the circuits using both approaches. The results, sorted by the simulation times for the cat-state method, can be seen in Figure~\ref{fig:hidden-shift-sorted}. In the most extreme case, for simulating the same circuit, the cat-state-based decomposition ran in less than half a second, while the BSS-based one took more than 1~h~25~m.

\begin{figure}[!htb]
\includegraphics[width=0.48\columnwidth]{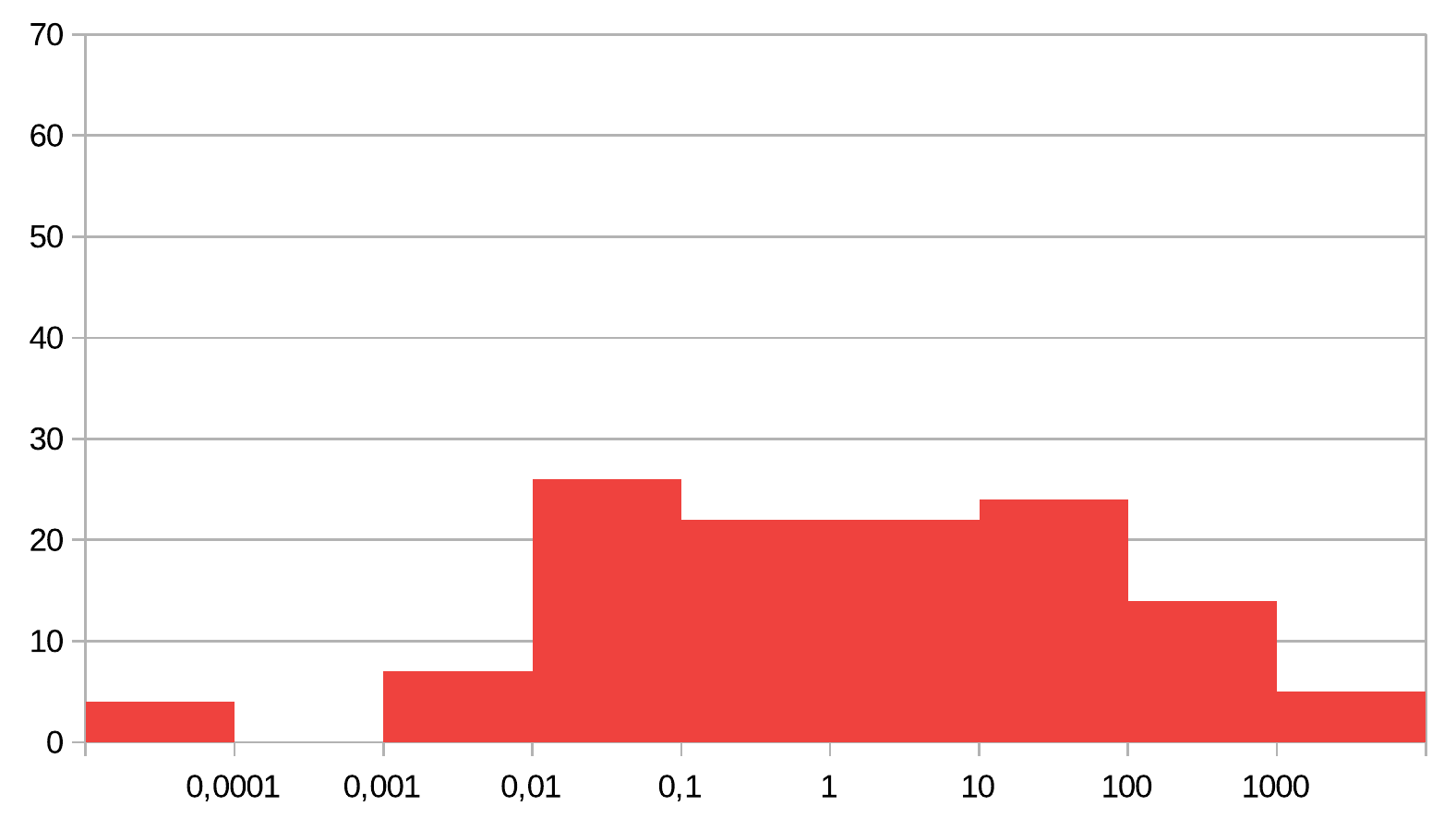} \quad
\includegraphics[width=0.48\columnwidth]{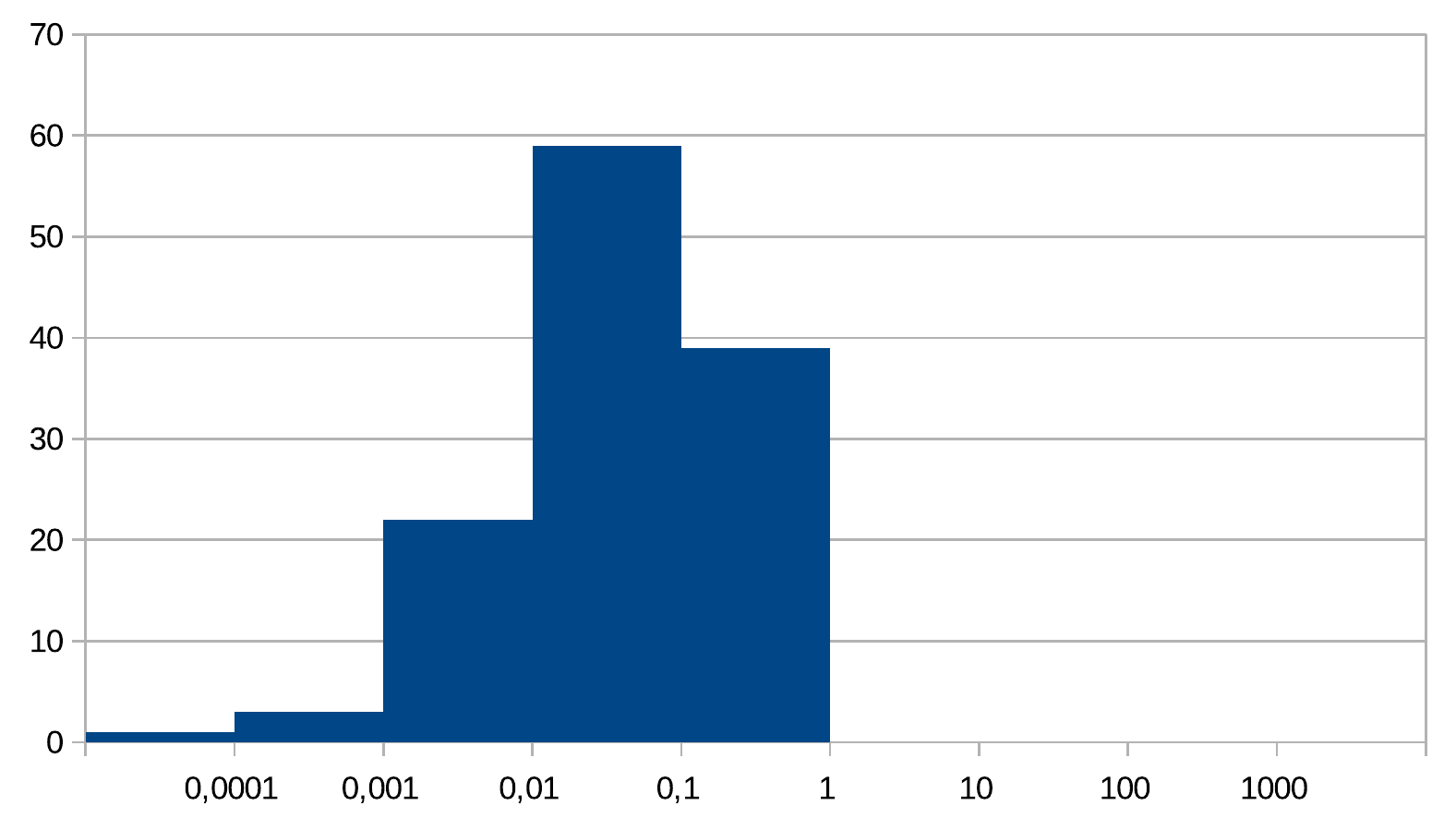}
\vspace{0.4cm}
\caption{Distributions of running times (in seconds) of simulating 125 different 20-qubit hidden-shift circuits with T-count 112 using the BSS decomposition (left) and the cat-state decomposition (right).}
\label{fig:hidden-shift-distributions}
\end{figure}

The distribution of running times for both approaches can be found in Figure~\ref{fig:hidden-shift-distributions}. As we are dealing with an inherently exponential process, it makes sense that perturbations have an exponential effect on the result. This is why we plotted the distributions on a logarithmic scale. In this same scale, it is possible to compute the variance of both distributions. We get a variance of $\sigma^2\simeq 0.523$ for the cat-state approach, and a variance of $\sigma^2\simeq 3.02$ for the BSS approach. This is strong evidence that the erratic behaviour is indeed dampened with our decomposition scheme, even when taking the exponential scaling into account.

\begin{figure}[!htb]
\centering
\includegraphics[width=1.0\columnwidth]{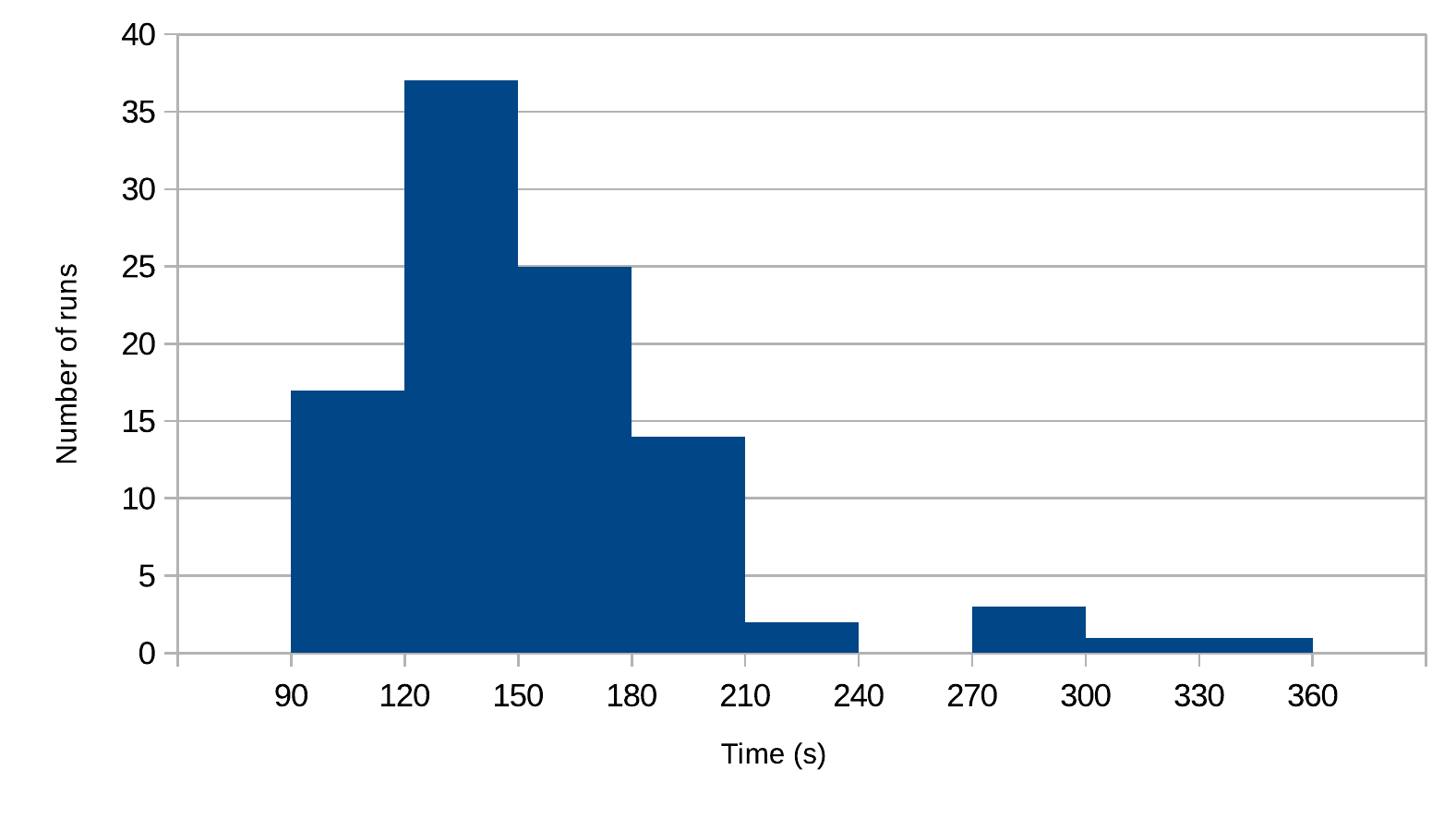}
% \vspace{-0.3cm}
\caption{The time distribution of simulating 100 random 50-qubit hidden-shift circuits with T-count 1400 using our new decompositions.}
\label{fig:hidden-shift-1400}
\end{figure}

%\begin{figure}[!htb]
%\includegraphics[width=\columnwidth]{cat-graphs/random-hidden-shift.eps}
%\caption{TODO}
%\end{figure}
%
%\begin{figure}[!htb]
%\includegraphics[width=\columnwidth]{cat-graphs/random-hidden-shift-only-cat.eps}
%\caption{TODO}
%\end{figure}
This erratic behaviour being lifted with the cat-state strategy, we were able to run tests without the need for a time limit, which Ref.~\cite{kissinger2021simulating} set at 5 minutes. We re-ran the experiment that performed strong simulation of 100 random 50-qubit hidden-shift circuits with T-count 1400 using our new decompositions. We observed that 98\% of the runs took less than 5 minutes, with the remaining circuits finishing within 6 minutes. This is to be compared with the BSS decomposition approach which in~\cite{kissinger2021simulating} obtained a 17\% success rate (with the 5 minutes limit) even though it ran on a dedicated server with better CPU and parallelisation. The running time distribution is given in Figure~\ref{fig:hidden-shift-1400}.

\section{Conclusion and Further Work}\label{sec:conclusion}

In this paper we built on the stabiliser rank quantum simulation method by combining it with a ZX-calculus simplification strategy and novel decompositions of entangled non-stabiliser states. We additionally introduced a new technique of \emph{partial} magic state decompositions that only reduce T-count instead of eliminating all at once, which achieves the stabiliser rank upper bound of $2^{0.396 t}$ found in~\cite{qassim2021improved} directly instead of just asymptotically.
Our benchmarks shows that our technique is in the best case several orders of magnitude faster than the strategy of~\cite{kissinger2021simulating}, which itself was already capable of simulating much larger circuits than the previous best. Additionally, our techniques seem to `smooth out' the erratic runtimes of~\cite{kissinger2021simulating} allowing for more consistent simulation times.

For future work it would be interesting to investigate whether other classes of entangled states with an even smaller stabiliser rank could be found and used for this kind of exact simulation. It is also worth investigating how decompositions of these states can be used in methods for \emph{approximate} simulations, like the stabiliser extent methods, or the norm estimation technique used in Ref.~\cite{Bravyi2019simulationofquantum}.

To sample from the random Clifford+T circuits we used the same technique as~\cite{kissinger2021simulating}, which required a technique that doubles the T-count in the worst case. It would be interesting to see if the recent technique of calculating marginals without doubling~\cite{bravyi2021simulate} would help improve run times.

\newpage
\bibliography{main}

\appendix

\newpage

\section{Plots}

\end{document}

%% file: preamble.tex
\usepackage[utf8]{inputenc}
\usepackage{amsthm,amsmath}
\usepackage{hyperref}

\usepackage[hyphenbreaks]{breakurl}
\usepackage[all]{hypcap}
\usepackage{stmaryrd}
\usepackage{graphicx}
\usepackage{keycommand}
\usepackage{multicol}
\usepackage{array}

\usepackage{tikzit}
\input{quantum.tikzstyles}
\input{quantum.tikzdefs}

\usepackage[all]{hypcap} % Make hyperlinks go to top of figures

\makeatletter
\newcommand\etc{etc\@ifnextchar.{}{.\@}\xspace}
\makeatother

%% file: quantum.tikzstyles
\tikzstyle{box}=[shape=rectangle, text height=1.5ex, text depth=0.25ex, yshift=0.5mm, fill=white, draw=black, minimum height=5mm, yshift=-0.5mm, minimum width=5mm, font={\small}]
\tikzstyle{gate}=[shape=rectangle, text height=1.5ex, text depth=0.25ex, yshift=0.5mm, fill=white, draw=black, minimum height=5mm, yshift=-0.5mm, minimum width=5mm, font={\small}, tikzit category=circuit]
\tikzstyle{big gate}=[shape=rectangle, text height=1.5ex, text depth=0.25ex, yshift=0.5mm, fill=white, draw=black, minimum height=10mm, yshift=-0.5mm, minimum width=5mm, font={\small}, tikzit category=circuit]
\tikzstyle{Z dot}=[inner sep=0mm, minimum size=2mm, shape=circle, draw=black, fill={rgb,255: red,221; green,255; blue,221}, tikzit category=zx]
\tikzstyle{Z phase dot}=[minimum size=5mm, font={\footnotesize\boldmath}, shape=rectangle, rounded corners=2mm, inner sep=0.2mm, outer sep=-2mm, scale=0.8, tikzit shape=circle, draw=black, fill={rgb,255: red,221; green,255; blue,221}, tikzit draw=blue, tikzit category=zx]
\tikzstyle{X dot}=[Z dot, shape=circle, draw=black, fill={rgb,255: red,255; green,136; blue,136}, tikzit category=zx]
\tikzstyle{X phase dot}=[Z phase dot, tikzit shape=circle, tikzit draw=blue, fill={rgb,255: red,255; green,136; blue,136}, font={\footnotesize\boldmath}, tikzit category=zx]
\tikzstyle{hadamard}=[fill=yellow, draw=black, shape=rectangle, inner sep=0.6mm, minimum height=1.5mm, minimum width=1.5mm, tikzit category=zx]
\tikzstyle{paulibox}=[fill={rgb,255: red,221; green,221; blue,255}, draw=black, shape=rectangle, inner sep=0.6mm, minimum height=5mm, minimum width=5mm, font={\footnotesize}, text height=1.5ex, text depth=0.25ex, tikzit category=zx]
\tikzstyle{vertex}=[inner sep=0mm, minimum size=1mm, shape=circle, draw=black, fill=black, tikzit category=misc]
\tikzstyle{vertex set}=[inner sep=0mm, minimum size=1mm, shape=circle, draw=black, fill=white, font={\footnotesize\boldmath}, tikzit category=misc]
\tikzstyle{small black dot}=[fill=black, draw=black, shape=circle, inner sep=0pt, minimum width=1.2mm, tikzit category=circuit]
\tikzstyle{cnot ctrl}=[fill=black, draw=black, shape=circle, inner sep=0pt, minimum width=1.2mm, tikzit category=circuit]
\tikzstyle{cnot targ}=[fill=white, draw=white, shape=circle, tikzit category=circuit, label={center:$\oplus$}, inner sep=0pt, minimum width=2.1mm, tikzit fill={rgb,255: red,102; green,204; blue,255}, tikzit draw=black]
\tikzstyle{ket}=[fill=white, draw=black, shape=regular polygon, regular polygon sides=3, regular polygon rotate=-30, scale=0.7, inner sep=1pt, tikzit category=circuit, tikzit shape=rectangle, tikzit fill=green]
\tikzstyle{bra}=[fill=white, draw=black, shape=regular polygon, regular polygon sides=3, regular polygon rotate=30, scale=0.7, inner sep=1pt, tikzit category=circuit, tikzit shape=rectangle, tikzit fill=red]
\tikzstyle{scalar}=[shape=rectangle, text height=1.5ex, text depth=0.25ex, yshift=0.5mm, fill=white, draw=black, minimum height=5mm, yshift=-0.5mm, minimum width=5mm, font={\small}]
\tikzstyle{clabel}=[fill=white, draw=none, shape=rectangle, tikzit fill={rgb,255: red,56; green,255; blue,242}, font={\footnotesize}, inner sep=1pt, tikzit category=labels]
\tikzstyle{empty diagram}=[draw={gray!40!white}, dashed, shape=rectangle, minimum width=1cm, minimum height=1cm, tikzit category=misc]

\tikzstyle{simple}=[-]
\tikzstyle{hadamard edge}=[-, dashed, dash pattern=on 2pt off 0.5pt, thick, draw={rgb,255: red,68; green,136; blue,255}]
\tikzstyle{box edge}=[-, dashed, dash pattern=on 2pt off 0.5pt, thick, draw={rgb,255: red,203; green,192; blue,225}]
\tikzstyle{brace edge}=[-, tikzit draw=blue, decorate, decoration={brace,amplitude=1mm,raise=-1mm}]
\tikzstyle{diredge}=[->]
\tikzstyle{double edge}=[-, double, shorten <=-1mm, shorten >=-1mm, double distance=2pt]
\tikzstyle{gray edge}=[-, {gray!60!white}]
\tikzstyle{pointer edge}=[->, very thick, gray]
\tikzstyle{boldedge}=[-, line width=1.6pt, shorten <=-0.17mm, shorten >=-0.17mm]

%% file: figures/Zsp-a.tikz
\begin{tikzpicture}
	\begin{pgfonlayer}{nodelayer}
		\node [style=Z phase dot] (0) at (0, 0) {$\alpha$};
		\node [style=none] (1) at (1.25, 1) {};
		\node [style=none] (2) at (-1.25, 1) {};
		\node [style=none] (3) at (-1.25, -1) {};
		\node [style=none] (4) at (1.25, -1) {};
		\node [style=none] (5) at (1.25, 0.5) {};
		\node [style=none] (6) at (-1.25, 0.5) {};
		\node [style=none, rotate=90] (7) at (-1, -0.25) {...};
		\node [style=none, rotate=90] (8) at (1, -0.25) {...};
	\end{pgfonlayer}
	\begin{pgfonlayer}{edgelayer}
		\draw [in=-141, out=0, looseness=0.75] (3.center) to (0);
		\draw [in=180, out=-39, looseness=0.75] (0) to (4.center);
		\draw [in=180, out=22, looseness=0.75] (0) to (5.center);
		\draw [in=180, out=39, looseness=0.75] (0) to (1.center);
		\draw [in=0, out=158, looseness=0.75] (0) to (6.center);
		\draw [in=141, out=0, looseness=0.75] (2.center) to (0);
	\end{pgfonlayer}
\end{tikzpicture}

%% file: figures/Xsp-a.tikz
\begin{tikzpicture}
	\begin{pgfonlayer}{nodelayer}
		\node [style=X phase dot] (0) at (0, 0) {$\alpha$};
		\node [style=none] (1) at (1.25, 1) {};
		\node [style=none] (2) at (-1.25, 1) {};
		\node [style=none] (3) at (-1.25, -1) {};
		\node [style=none] (4) at (1.25, -1) {};
		\node [style=none] (5) at (1.25, 0.5) {};
		\node [style=none] (6) at (-1.25, 0.5) {};
		\node [style=none, rotate=90] (7) at (-1, -0.25) {...};
		\node [style=none, rotate=90] (8) at (1, -0.25) {...};
	\end{pgfonlayer}
	\begin{pgfonlayer}{edgelayer}
		\draw [in=-141, out=0, looseness=0.75] (3.center) to (0);
		\draw [in=180, out=-39, looseness=0.75] (0) to (4.center);
		\draw [in=180, out=22, looseness=0.75] (0) to (5.center);
		\draw [in=180, out=39, looseness=0.75] (0) to (1.center);
		\draw [in=0, out=158, looseness=0.75] (0) to (6.center);
		\draw [in=141, out=0, looseness=0.75] (2.center) to (0);
	\end{pgfonlayer}
\end{tikzpicture}

%% file: figures/h-alone.tikz
\begin{tikzpicture}
	\begin{pgfonlayer}{nodelayer}
		\node [style=none] (0) at (-0.25, 0) {};
		\node [style=none] (1) at (1.75, 0) {};
		\node [style=hadamard] (2) at (0.75, 0) {};
	\end{pgfonlayer}
	\begin{pgfonlayer}{edgelayer}
		\draw (0.center) to (2);
		\draw (2) to (1.center);
	\end{pgfonlayer}
\end{tikzpicture}

%% file: figures/id.tikz
\begin{tikzpicture}
	\begin{pgfonlayer}{nodelayer}
		\node [style=none] (0) at (-1, 0) {};
		\node [style=none] (1) at (0.5, 0) {};
	\end{pgfonlayer}
	\begin{pgfonlayer}{edgelayer}
		\draw (0.center) to (1.center);
	\end{pgfonlayer}
\end{tikzpicture}

%% file: figures/swap.tikz
\begin{tikzpicture}
	\begin{pgfonlayer}{nodelayer}
		\node [style=none] (3) at (-1.75, -0.5) {};
		\node [style=none] (4) at (-0.75, 0.5) {};
		\node [style=none] (5) at (-0.75, -0.5) {};
		\node [style=none] (6) at (-1.75, 0.5) {};
	\end{pgfonlayer}
	\begin{pgfonlayer}{edgelayer}
		\draw [in=180, out=0] (6.center) to (5.center);
		\draw [in=360, out=180] (4.center) to (3.center);
	\end{pgfonlayer}
\end{tikzpicture}

%% file: figures/cap.tikz
\begin{tikzpicture}
	\begin{pgfonlayer}{nodelayer}
		\node [style=none] (0) at (0.5, 0.5) {};
		\node [style=none] (1) at (0.5, -0.5) {};
	\end{pgfonlayer}
	\begin{pgfonlayer}{edgelayer}
		\draw [bend left=90, looseness=1.75] (1.center) to (0.center);
	\end{pgfonlayer}
\end{tikzpicture}

%% file: figures/cup.tikz
\begin{tikzpicture}
	\begin{pgfonlayer}{nodelayer}
		\node [style=none] (0) at (0.5, 0.5) {};
		\node [style=none] (1) at (0.5, -0.5) {};
	\end{pgfonlayer}
	\begin{pgfonlayer}{edgelayer}
		\draw [bend right=90, looseness=1.75] (1.center) to (0.center);
	\end{pgfonlayer}
\end{tikzpicture}

%% file: figures/D2.tikz
\begin{tikzpicture}
	\begin{pgfonlayer}{nodelayer}
		\node [style=none] (0) at (0.5, 0.5) {};
		\node [style=none] (1) at (1.25, 0.5) {};
		\node [style=none] (2) at (1.25, 0.75) {};
		\node [style=none] (3) at (0.5, -0.5) {};
		\node [style=none] (4) at (1.25, -0.5) {};
		\node [style=none] (5) at (1.25, -0.75) {};
		\node [style=none, rotate=90] (6) at (0.75, 0) {...};
		\node [style=none] (7) at (3, 0.5) {};
		\node [style=none] (8) at (2.25, 0.5) {};
		\node [style=none] (9) at (2.25, 0.75) {};
		\node [style=none] (10) at (3, -0.5) {};
		\node [style=none] (11) at (2.25, -0.5) {};
		\node [style=none] (12) at (2.25, -0.75) {};
		\node [style=none, rotate=90] (13) at (2.75, 0) {...};
		\node [style=none] (14) at (1.75, 0) {$D_2$};
	\end{pgfonlayer}
	\begin{pgfonlayer}{edgelayer}
		\draw (0.center) to (1.center);
		\draw (3.center) to (4.center);
		\draw (2.center) to (5.center);
		\draw (7.center) to (8.center);
		\draw (10.center) to (11.center);
		\draw (9.center) to (12.center);
		\draw (2.center) to (9.center);
		\draw (12.center) to (5.center);
	\end{pgfonlayer}
\end{tikzpicture}

%% file: figures/D1.tikz
\begin{tikzpicture}
	\begin{pgfonlayer}{nodelayer}
		\node [style=none] (0) at (-0.75, 0.5) {};
		\node [style=none] (1) at (0, 0.5) {};
		\node [style=none] (2) at (0, 0.75) {};
		\node [style=none] (3) at (-0.75, -0.5) {};
		\node [style=none] (4) at (0, -0.5) {};
		\node [style=none] (5) at (0, -0.75) {};
		\node [style=none, rotate=90] (6) at (-0.5, 0) {...};
		\node [style=none] (7) at (1.75, 0.5) {};
		\node [style=none] (8) at (1, 0.5) {};
		\node [style=none] (9) at (1, 0.75) {};
		\node [style=none] (10) at (1.75, -0.5) {};
		\node [style=none] (11) at (1, -0.5) {};
		\node [style=none] (12) at (1, -0.75) {};
		\node [style=none, rotate=90] (13) at (1.5, 0) {...};
		\node [style=none] (14) at (0.5, 0) {$D_1$};
	\end{pgfonlayer}
	\begin{pgfonlayer}{edgelayer}
		\draw (0.center) to (1.center);
		\draw (3.center) to (4.center);
		\draw (2.center) to (5.center);
		\draw (7.center) to (8.center);
		\draw (10.center) to (11.center);
		\draw (9.center) to (12.center);
		\draw (2.center) to (9.center);
		\draw (12.center) to (5.center);
	\end{pgfonlayer}
\end{tikzpicture}

%% file: figures/cnot-aux.tikz
\begin{tikzpicture}
	\begin{pgfonlayer}{nodelayer}
		\node [style=Z dot] (0) at (1.5, 0.5) {};
		\node [style=none] (1) at (3, 0.5) {};
		\node [style=X dot] (2) at (2, -0.5) {};
		\node [style=none] (3) at (0.5, 0.5) {};
		\node [style=none] (4) at (3, -0.5) {};
		\node [style=none] (5) at (0.5, -0.5) {};
	\end{pgfonlayer}
	\begin{pgfonlayer}{edgelayer}
		\draw (0) to (2);
		\draw (0) to (3.center);
		\draw (0) to (1.center);
		\draw (5.center) to (2);
		\draw (2) to (4.center);
	\end{pgfonlayer}
\end{tikzpicture}

%% file: figures/Z-a.tikz
\begin{tikzpicture}
	\begin{pgfonlayer}{nodelayer}
		\node [style=Z phase dot] (0) at (0, 0) {$\alpha$};
		\node [style=none] (1) at (-1.25, 0) {};
		\node [style=none] (2) at (1.25, 0) {};
	\end{pgfonlayer}
	\begin{pgfonlayer}{edgelayer}
		\draw (1.center) to (0);
		\draw (0) to (2.center);
	\end{pgfonlayer}
\end{tikzpicture}

%% file: figures/spider-rule.tikz
\begin{tikzpicture}
	\begin{pgfonlayer}{nodelayer}
		\node [style=Z phase dot] (0) at (-2.25, -0.75) {$\beta$};
		\node [style=none] (1) at (-1.25, -0.25) {};
		\node [style=none] (2) at (-4, -0.25) {};
		\node [style=none] (3) at (-4, -1.5) {};
		\node [style=none] (4) at (-1.25, -1.5) {};
		\node [style=none] (5) at (-1.25, -0.5) {};
		\node [style=none] (6) at (-4, -0.5) {};
		\node [style=none, rotate=90] (7) at (-4.25, -1) {...};
		\node [style=none, rotate=90] (8) at (-1.25, -1) {...};
		\node [style=none] (9) at (-4.75, 1) {};
		\node [style=Z phase dot] (10) at (-3.75, 0.75) {$\alpha$};
		\node [style=none] (11) at (-1.75, 1.25) {};
		\node [style=none] (12) at (-4.75, 1.25) {};
		\node [style=none] (13) at (-1.75, 0) {};
		\node [style=none, rotate=90] (14) at (-1.75, 0.5) {...};
		\node [style=none] (15) at (-1.75, 1) {};
		\node [style=none] (16) at (-4.75, 0) {};
		\node [style=none, rotate=90] (17) at (-4.5, 0.5) {...};
		\node [style=none] (18) at (-1.25, 0) {};
		\node [style=none] (19) at (-1.25, 1.25) {};
		\node [style=none] (20) at (-1.25, 1) {};
		\node [style=none] (21) at (-4.75, -0.5) {};
		\node [style=none] (22) at (-4.75, -1.5) {};
		\node [style=none] (23) at (-4.75, -0.25) {};
		\node [style=none] (24) at (0, 0) {$=$};
		\node [style=none, rotate=45] (25) at (-3, 0) {...};
		\node [style=none] (26) at (1.25, 1.25) {};
		\node [style=none] (27) at (4.75, 0.75) {};
		\node [style=none, rotate=90] (28) at (1.75, -0.25) {...};
		\node [style=none] (29) at (4.75, -1.5) {};
		\node [style=none, rotate=90] (30) at (4.5, -0.25) {...};
		\node [style=none] (31) at (1.25, 0.75) {};
		\node [style=Z phase dot] (32) at (3, 0) {$\ \alpha\!+\!\beta\ $};
		\node [style=none] (33) at (4.75, 1.25) {};
		\node [style=none] (34) at (1.25, -1.5) {};
	\end{pgfonlayer}
	\begin{pgfonlayer}{edgelayer}
		\draw [style=simple, in=-141, out=0, looseness=0.75] (3.center) to (0);
		\draw [style=simple, in=180, out=-39, looseness=0.75] (0) to (4.center);
		\draw [style=simple, in=180, out=22, looseness=0.75] (0) to (5.center);
		\draw [style=simple, in=180, out=39, looseness=0.75] (0) to (1.center);
		\draw [style=simple, in=0, out=180] (0) to (6.center);
		\draw [style=simple, in=165, out=0] (2.center) to (0);
		\draw [style=simple, in=-141, out=0, looseness=0.75] (16.center) to (10);
		\draw [style=simple, in=180, out=-15] (10) to (13.center);
		\draw [style=simple, in=180, out=22] (10) to (15.center);
		\draw [style=simple, in=180, out=39, looseness=0.75] (10) to (11.center);
		\draw [style=simple, in=0, out=158, looseness=0.75] (10) to (9.center);
		\draw [style=simple, in=141, out=0, looseness=0.75] (12.center) to (10);
		\draw [style=simple, bend left] (10) to (0);
		\draw [style=simple] (11.center) to (19.center);
		\draw [style=simple] (15.center) to (20.center);
		\draw [style=simple] (13.center) to (18.center);
		\draw [style=simple] (23.center) to (2.center);
		\draw [style=simple] (21.center) to (6.center);
		\draw [style=simple] (22.center) to (3.center);
		\draw [style=simple, bend left] (0) to (10);
		\draw [style=simple, in=-120, out=0] (34.center) to (32);
		\draw [style=simple, in=180, out=-60] (32) to (29.center);
		\draw [style=simple, in=180, out=45, looseness=0.75] (32) to (27.center);
		\draw [style=simple, in=180, out=60] (32) to (33.center);
		\draw [style=simple, in=0, out=135, looseness=0.75] (32) to (31.center);
		\draw [style=simple, in=120, out=0] (26.center) to (32);
	\end{pgfonlayer}
\end{tikzpicture}

%% file: figures/h-involution-rule.tikz
\begin{tikzpicture}
	\begin{pgfonlayer}{nodelayer}
		\node [style=none] (0) at (0, 0) {$=$};
		\node [style=none] (1) at (2, 0) {};
		\node [style=none] (2) at (-1, 0) {};
		\node [style=none] (3) at (-2.5, 0) {};
		\node [style=none] (4) at (1, 0) {};
		\node [style=hadamard] (5) at (-2, 0) {};
		\node [style=hadamard] (6) at (-1.5, 0) {};
	\end{pgfonlayer}
	\begin{pgfonlayer}{edgelayer}
		\draw (3.center) to (2.center);
		\draw (4.center) to (1.center);
	\end{pgfonlayer}
\end{tikzpicture}

%% file: figures/0-rotation.tikz
\begin{tikzpicture}
	\begin{pgfonlayer}{nodelayer}
		\node [style=Z dot] (0) at (-1.75, 0) {};
		\node [style=none] (1) at (-2.5, 0) {};
		\node [style=none] (2) at (-1, 0) {};
		\node [style=none] (3) at (1, 0) {};
		\node [style=none] (4) at (2, 0) {};
		\node [style=none] (5) at (0, 0) {$=$};
	\end{pgfonlayer}
	\begin{pgfonlayer}{edgelayer}
		\draw (1.center) to (2.center);
		\draw (3.center) to (4.center);
	\end{pgfonlayer}
\end{tikzpicture}

%% file: figures/colour-change-rule.tikz
\begin{tikzpicture}
	\begin{pgfonlayer}{nodelayer}
		\node [style=hadamard] (0) at (1, 1.25) {};
		\node [style=none, rotate=90] (3) at (1, -0.25) {...};
		\node [style=none] (4) at (-2.25, 0) {$=$};
		\node [style=hadamard] (5) at (1, -1) {};
		\node [style=none] (8) at (1.5, 1.25) {};
		\node [style=none] (10) at (1.5, 0.5) {};
		\node [style=hadamard] (11) at (1, 0.5) {};
		\node [style=none] (13) at (1.5, -1) {};
		\node [style=X phase dot] (17) at (0, 0) {$\alpha$};
		\node [style=none] (18) at (-1.5, -1) {};
		\node [style=none, rotate=90] (19) at (-1.25, -0.25) {...};
		\node [style=none] (20) at (-1.5, 1.25) {};
		\node [style=none] (21) at (-1.5, 0.5) {};
		\node [style=hadamard] (22) at (-5.5, 1.25) {};
		\node [style=none, rotate=90] (23) at (-5.5, -0.25) {...};
		\node [style=hadamard] (24) at (-5.5, -1) {};
		\node [style=none] (25) at (-6, 1.25) {};
		\node [style=none] (26) at (-6, 0.5) {};
		\node [style=hadamard] (27) at (-5.5, 0.5) {};
		\node [style=none] (28) at (-6, -1) {};
		\node [style=Z phase dot] (29) at (-4.5, 0) {$\alpha$};
		\node [style=none] (30) at (-3, -1) {};
		\node [style=none, rotate=90] (31) at (-3.25, -0.25) {...};
		\node [style=none] (32) at (-3, 1.25) {};
		\node [style=none] (33) at (-3, 0.5) {};
	\end{pgfonlayer}
	\begin{pgfonlayer}{edgelayer}
		\draw [style=simple] (8.center) to (0);
		\draw [style=simple] (10.center) to (11);
		\draw [style=simple] (13.center) to (5);
		\draw [style=simple, in=0, out=-120, looseness=0.75] (17) to (18.center);
		\draw [style=simple, in=0, out=135, looseness=0.75] (17) to (21.center);
		\draw [style=simple, in=0, out=105, looseness=0.75] (17) to (20.center);
		\draw [style=simple, in=-60, out=180, looseness=0.75] (5) to (17);
		\draw [style=simple, in=180, out=45, looseness=0.75] (17) to (11);
		\draw [style=simple, in=75, out=180, looseness=0.75] (0) to (17);
		\draw [style=simple] (25.center) to (22);
		\draw [style=simple] (26.center) to (27);
		\draw [style=simple] (28.center) to (24);
		\draw [style=simple, in=180, out=-60, looseness=0.75] (29) to (30.center);
		\draw [style=simple, in=180, out=45, looseness=0.75] (29) to (33.center);
		\draw [style=simple, in=180, out=75, looseness=0.75] (29) to (32.center);
		\draw [style=simple, in=-120, out=0, looseness=0.75] (24) to (29);
		\draw [style=simple, in=0, out=135, looseness=0.75] (29) to (27);
		\draw [style=simple, in=105, out=0, looseness=0.75] (22) to (29);
	\end{pgfonlayer}
\end{tikzpicture}

%% file: figures/neighbouring-cliffords.tikz
\begin{tikzpicture}
	\begin{pgfonlayer}{nodelayer}
		\node [style=none] (5) at (-2.25, -0.75) {};
		\node [style=none] (6) at (-1, 1) {};
		\node [style=none] (7) at (2.25, -0.75) {};
		\node [style=none] (8) at (2.25, 1) {};
		\node [style=Z phase dot] (9) at (-1.5, 0) {$\ k\frac\pi2\ $};
		\node [style=Z phase dot] (10) at (1.5, 0) {$\ \ell\frac\pi2\ $};
		\node [style=none] (11) at (-2.25, 1) {};
		\node [style=none] (12) at (0.5, -1) {};
	\end{pgfonlayer}
	\begin{pgfonlayer}{edgelayer}
		\draw [style=hadamard edge] (8.center) to (10);
		\draw (10) to (7.center);
		\draw (5.center) to (9);
		\draw [style=hadamard edge] (9) to (6.center);
		\draw [style=hadamard edge] (12.center) to (10);
		\draw (11.center) to (9);
		\draw [style=hadamard edge] (9) to (10);
	\end{pgfonlayer}
\end{tikzpicture}

%% file: figures/binary-clifford.tikz
\begin{tikzpicture}
	\begin{pgfonlayer}{nodelayer}
		\node [style=Z phase dot] (0) at (0, 0) {$\ k\frac\pi2\ $};
		\node [style=none] (1) at (-1.5, 0) {};
		\node [style=none] (2) at (1.5, 0) {};
	\end{pgfonlayer}
	\begin{pgfonlayer}{edgelayer}
		\draw [style=hadamard edge] (1.center) to (2.center);
	\end{pgfonlayer}
\end{tikzpicture}

%% file: figures/unary-clifford.tikz
\begin{tikzpicture}
	\begin{pgfonlayer}{nodelayer}
		\node [style=Z phase dot] (0) at (0.5, 0) {$\ k\frac\pi2\ $};
		\node [style=none] (1) at (2, 0) {};
		\node [style=none] (2) at (2, 0) {};
	\end{pgfonlayer}
	\begin{pgfonlayer}{edgelayer}
		\draw [style=hadamard edge] (0) to (2.center);
	\end{pgfonlayer}
\end{tikzpicture}

%% file: figures/internal-S-spider.tikz
\begin{tikzpicture}
	\begin{pgfonlayer}{nodelayer}
		\node [style=none] (0) at (-2, -1) {};
		\node [style=none] (1) at (2, 1) {};
		\node [style=Z phase dot] (2) at (0, 0) {$\ (2k{+}1)\frac\pi2\ $};
		\node [style=none] (3) at (-2, 1) {};
		\node [style=none] (4) at (2, -1) {};
		\node [style=none, rotate=90] (5) at (1.75, 0) {...};
		\node [style=none, rotate=90] (6) at (-1.75, 0) {...};
	\end{pgfonlayer}
	\begin{pgfonlayer}{edgelayer}
		\draw [style=hadamard edge] (0.center) to (2);
		\draw [style=hadamard edge] (2) to (1.center);
		\draw [style=hadamard edge] (3.center) to (2);
		\draw [style=hadamard edge] (4.center) to (2);
	\end{pgfonlayer}
\end{tikzpicture}

%% file: figures/magic-state.tikz
\begin{tikzpicture}[scale=2]
	\begin{pgfonlayer}{nodelayer}
		\node [style=Z phase dot] (0) at (0, -0.012) {$\frac\pi4$};
		\node [style=none] (1) at (0.525, -0.013) {};
	\end{pgfonlayer}
	\begin{pgfonlayer}{edgelayer}
		\draw (1.center) to (0);
	\end{pgfonlayer}
\end{tikzpicture}

%% file: figures/T-spider-to-magic-state.tikz
\begin{tikzpicture}[scale=2.5]
	\begin{pgfonlayer}{nodelayer}
		\node [style=Z phase dot] (0) at (-1.75, 0) {$~(2k{+}1)\frac\pi4~$};
		\node [style=none] (1) at (-0.75, 0.25) {};
		\node [style=none] (2) at (-2.75, -0.25) {};
		\node [style=none] (3) at (-0.75, -0.25) {};
		\node [style=none] (4) at (-2.75, 0.25) {};
		\node [style=none] (8) at (0, 0) {$=$};
		\node [style=Z phase dot] (9) at (1.5, 0) {$~k\frac\pi2~$};
		\node [style=none] (10) at (0.75, -0.25) {};
		\node [style=none] (11) at (2.25, 0.25) {};
		\node [style=none] (12) at (0.75, 0.25) {};
		\node [style=none] (13) at (2.25, -0.25) {};
		\node [style=Z phase dot] (17) at (1, 0.5) {$\frac\pi4$};
		\node [style=none, rotate=90] (18) at (-2.625, 0) {...};
		\node [style=none, rotate=90] (19) at (-0.875, 0) {...};
		\node [style=none, rotate=90] (20) at (0.875, 0) {...};
		\node [style=none, rotate=90] (21) at (2.125, 0) {...};
	\end{pgfonlayer}
	\begin{pgfonlayer}{edgelayer}
		\draw (1.center) to (2.center);
		\draw (3.center) to (4.center);
		\draw (10.center) to (11.center);
		\draw (12.center) to (13.center);
		\draw [in=-15, out=120] (9) to (17);
	\end{pgfonlayer}
\end{tikzpicture}

%% file: figures/ket-0-GL.tikz
\begin{tikzpicture}
	\begin{pgfonlayer}{nodelayer}
		\node [style=Z dot] (0) at (-1, 0) {};
		\node [style=Z dot] (1) at (0, 0) {};
		\node [style=none] (2) at (1, 0) {};
	\end{pgfonlayer}
	\begin{pgfonlayer}{edgelayer}
		\draw (1) to (2.center);
		\draw [style=hadamard edge] (0) to (1);
	\end{pgfonlayer}
\end{tikzpicture}

%% file: figures/ket-1-GL.tikz
\begin{tikzpicture}
	\begin{pgfonlayer}{nodelayer}
		\node [style=Z phase dot] (0) at (-1, 0) {$\pi$};
		\node [style=Z dot] (1) at (0.25, 0) {};
		\node [style=none] (2) at (1.25, 0) {};
	\end{pgfonlayer}
	\begin{pgfonlayer}{edgelayer}
		\draw (1) to (2.center);
		\draw [style=hadamard edge] (0) to (1);
	\end{pgfonlayer}
\end{tikzpicture}

%% file: figures/2-T-decomposition.tikz
\begin{tikzpicture}
	\begin{pgfonlayer}{nodelayer}
		\node [style=Z phase dot] (0) at (-1.5, 0.488) {$\frac\pi4$};
		\node [style=none] (1) at (-0.475, 0.487) {};
		\node [style=Z phase dot] (2) at (-1.475, -0.512) {$\frac\pi4$};
		\node [style=none] (3) at (-0.45, -0.513) {};
		\node [style=none] (4) at (0.5, 0) {$=$};
		\node [style=Z phase dot] (5) at (1.75, -0.012) {$\frac\pi2$};
		\node [style=none] (6) at (2.775, 0.487) {};
		\node [style=none] (8) at (2.8, -0.513) {};
		\node [style=none] (9) at (4, 0) {$+\ e^{i\frac\pi4}$};
		\node [style=Z phase dot] (10) at (5.5, 0) {$\pi$};
		\node [style=Z dot] (11) at (6.75, 0.5) {};
		\node [style=Z dot] (12) at (6.75, -0.5) {};
		\node [style=none] (13) at (7.5, 0.5) {};
		\node [style=none] (14) at (7.5, -0.5) {};
	\end{pgfonlayer}
	\begin{pgfonlayer}{edgelayer}
		\draw (1.center) to (0);
		\draw (3.center) to (2);
		\draw [bend right] (6.center) to (5);
		\draw [bend right] (5) to (8.center);
		\draw (13.center) to (11);
		\draw (14.center) to (12);
		\draw [style=hadamard edge] (10) to (12);
		\draw [style=hadamard edge] (10) to (11);
	\end{pgfonlayer}
\end{tikzpicture}

%% file: figures/magic-state-bot.tikz
\begin{tikzpicture}[scale=2]
	\begin{pgfonlayer}{nodelayer}
		\node [style=Z phase dot] (0) at (0, -0.012) {$\frac{5\pi}4$};
		\node [style=none] (1) at (0.425, -0.013) {};
	\end{pgfonlayer}
	\begin{pgfonlayer}{edgelayer}
		\draw (1.center) to (0);
	\end{pgfonlayer}
\end{tikzpicture}

%% file: figures/cat-state.tikz
\begin{tikzpicture}[scale=2]
	\begin{pgfonlayer}{nodelayer}
		\node [style=Z phase dot] (0) at (-0.113, 0.65) {$\frac\pi4$};
		\node [style=none] (1) at (0.238, 0.65) {};
		\node [style=Z phase dot] (3) at (-0.113, 0) {$\frac\pi4$};
		\node [style=none] (4) at (0.238, 0) {};
		\node [style=Z phase dot] (6) at (-0.113, -1.15) {$\frac\pi4$};
		\node [style=none] (7) at (0.238, -1.15) {};
		\node [style=Z dot] (9) at (-0.75, -0.25) {};
		\node [style=none] (10) at (-0.125, -0.475) {\vdots};
	\end{pgfonlayer}
	\begin{pgfonlayer}{edgelayer}
		\draw (1.center) to (0);
		\draw (4.center) to (3);
		\draw (7.center) to (6);
		\draw [style=hadamard edge, in=-165, out=90] (9) to (0);
		\draw [style=hadamard edge, in=-180, out=45] (9) to (3);
		\draw [style=hadamard edge, in=165, out=-90] (9) to (6);
	\end{pgfonlayer}
\end{tikzpicture}

%% file: figures/cat-6-decomp.tikz
\begin{tikzpicture}[xscale=1.7, yscale=2]
	\begin{pgfonlayer}{nodelayer}
		\node [style=Z phase dot] (0) at (-1.363, -1.425) {$\frac\pi4$};
		\node [style=none] (1) at (-0.762, -1.425) {};
		\node [style=Z phase dot] (3) at (-1.363, -0.825) {$\frac\pi4$};
		\node [style=none] (4) at (-0.762, -0.825) {};
		\node [style=Z dot] (9) at (-2, 0.05) {};
		\node [style=Z phase dot] (10) at (-1.363, -0.225) {$\frac\pi4$};
		\node [style=none] (11) at (-0.762, -0.225) {};
		\node [style=Z phase dot] (13) at (-1.363, 0.35) {$\frac\pi4$};
		\node [style=none] (14) at (-0.762, 0.35) {};
		\node [style=Z phase dot] (16) at (-1.363, 0.95) {$\frac\pi4$};
		\node [style=none] (17) at (-0.762, 0.95) {};
		\node [style=Z phase dot] (19) at (-1.363, 1.55) {$\frac\pi4$};
		\node [style=none] (20) at (-0.762, 1.55) {};
		\node [style=none] (22) at (0.25, 0) {$=$};
		\node [style=none] (23) at (2.512, -1.45) {};
		\node [style=none] (26) at (2.512, -0.85) {};
		\node [style=Z phase dot] (29) at (1.625, 0.025) {$~-\frac\pi2~$};
		\node [style=none] (30) at (2.512, -0.25) {};
		\node [style=none] (33) at (2.512, 0.325) {};
		\node [style=none] (36) at (2.512, 0.925) {};
		\node [style=none] (39) at (2.512, 1.525) {};
		\node [style=none] (41) at (3.75, 0) {$+~\frac{i e^{i\pi/4}}{\sqrt2}$};
		\node [style=Z dot] (42) at (5.512, -1.425) {};
		\node [style=Z dot] (43) at (5.512, -0.825) {};
		\node [style=Z dot] (44) at (4.875, 0.05) {};
		\node [style=Z dot] (45) at (5.512, -0.225) {};
		\node [style=Z dot] (46) at (5.512, 0.35) {};
		\node [style=Z dot] (47) at (5.512, 0.95) {};
		\node [style=Z dot] (48) at (5.512, 1.55) {};
		\node [style=none] (49) at (6.75, 0) {$-\ \frac{e^{i\pi/4}}{\sqrt{2}}$};
		\node [style=none] (68) at (5.887, -1.425) {};
		\node [style=none] (69) at (5.887, -0.825) {};
		\node [style=none] (70) at (5.887, -0.225) {};
		\node [style=none] (71) at (5.887, 0.35) {};
		\node [style=none] (72) at (5.887, 0.95) {};
		\node [style=none] (73) at (5.887, 1.55) {};
		\node [style=none] (74) at (0.75, 0) {$\frac12$};
		\node [style=Z phase dot] (75) at (8.512, -1.425) {$\frac\pi2$};
		\node [style=Z phase dot] (76) at (8.512, -0.825) {$\frac\pi2$};
		\node [style=Z dot] (77) at (7.625, 0.05) {};
		\node [style=Z phase dot] (78) at (8.512, -0.225) {$\frac\pi2$};
		\node [style=Z phase dot] (79) at (8.512, 0.35) {$\frac\pi2$};
		\node [style=Z phase dot] (80) at (8.512, 0.95) {$\frac\pi2$};
		\node [style=Z phase dot] (81) at (8.512, 1.55) {$\frac\pi2$};
		\node [style=none] (82) at (9.012, -1.425) {};
		\node [style=none] (83) at (9.012, -0.825) {};
		\node [style=none] (84) at (9.012, -0.225) {};
		\node [style=none] (85) at (9.012, 0.35) {};
		\node [style=none] (86) at (9.012, 0.95) {};
		\node [style=none] (87) at (9.012, 1.55) {};
	\end{pgfonlayer}
	\begin{pgfonlayer}{edgelayer}
		\draw (1.center) to (0);
		\draw (4.center) to (3);
		\draw [style=hadamard edge, in=165, out=-90] (9) to (0);
		\draw [style=hadamard edge, in=165, out=-75] (9) to (3);
		\draw (11.center) to (10);
		\draw (14.center) to (13);
		\draw (17.center) to (16);
		\draw (20.center) to (19);
		\draw [style=hadamard edge, in=180, out=-45] (9) to (10);
		\draw [style=hadamard edge, in=45, out=180] (13) to (9);
		\draw [style=hadamard edge, in=-165, out=75] (9) to (16);
		\draw [style=hadamard edge, in=-165, out=90] (9) to (19);
		\draw [in=165, out=-90] (29) to (23.center);
		\draw [in=165, out=-75] (29) to (26.center);
		\draw [in=180, out=-45] (29) to (30.center);
		\draw [in=45, out=180] (33.center) to (29);
		\draw [in=-165, out=75] (29) to (36.center);
		\draw [in=-165, out=90] (29) to (39.center);
		\draw [style=hadamard edge, in=165, out=-90] (44) to (42);
		\draw [style=hadamard edge, in=165, out=-75] (44) to (43);
		\draw [style=hadamard edge, in=180, out=-45] (44) to (45);
		\draw [style=hadamard edge, in=45, out=180] (46) to (44);
		\draw [style=hadamard edge, in=-165, out=75] (44) to (47);
		\draw [style=hadamard edge, in=-165, out=90] (44) to (48);
		\draw (42) to (68.center);
		\draw (43) to (69.center);
		\draw (45) to (70.center);
		\draw (46) to (71.center);
		\draw (47) to (72.center);
		\draw (48) to (73.center);
		\draw [style=hadamard edge, in=165, out=-90] (77) to (75);
		\draw [style=hadamard edge, in=165, out=-75] (77) to (76);
		\draw [style=hadamard edge, in=180, out=-45] (77) to (78);
		\draw [style=hadamard edge, in=45, out=180] (79) to (77);
		\draw [style=hadamard edge, in=-165, out=75] (77) to (80);
		\draw [style=hadamard edge, in=-165, out=90] (77) to (81);
		\draw (75) to (82.center);
		\draw (76) to (83.center);
		\draw (78) to (84.center);
		\draw (79) to (85.center);
		\draw (80) to (86.center);
		\draw (81) to (87.center);
	\end{pgfonlayer}
\end{tikzpicture}

%% file: figures/cat-4-decomp.tikz
\begin{tikzpicture}[scale=2]
	\begin{pgfonlayer}{nodelayer}
		\node [style=Z phase dot] (3) at (-1.113, -0.825) {$\frac\pi4$};
		\node [style=none] (4) at (-0.762, -0.825) {};
		\node [style=Z dot] (6) at (-1.75, 0.05) {};
		\node [style=Z phase dot] (7) at (-1.113, -0.225) {$\frac\pi4$};
		\node [style=none] (8) at (-0.762, -0.225) {};
		\node [style=Z phase dot] (10) at (-1.113, 0.35) {$\frac\pi4$};
		\node [style=none] (11) at (-0.762, 0.35) {};
		\node [style=Z phase dot] (13) at (-1.113, 0.95) {$\frac\pi4$};
		\node [style=none] (14) at (-0.762, 0.95) {};
		\node [style=none] (19) at (0, 0) {$=$};
		\node [style=none] (21) at (2.637, -0.875) {};
		\node [style=Z phase dot] (22) at (1.75, 0) {$~-\frac\pi2~$};
		\node [style=none] (23) at (2.637, -0.275) {};
		\node [style=none] (24) at (2.637, 0.3) {};
		\node [style=none] (25) at (2.637, 0.9) {};
		\node [style=none] (28) at (3.2, 0) {$+~i$};
		\node [style=Z dot] (30) at (4.387, -0.875) {};
		\node [style=Z dot] (31) at (3.75, 0) {};
		\node [style=Z dot] (32) at (4.387, -0.275) {};
		\node [style=Z dot] (33) at (4.387, 0.3) {};
		\node [style=Z dot] (34) at (4.387, 0.9) {};
		\node [style=none] (35) at (4.637, -0.875) {};
		\node [style=none] (36) at (4.637, -0.275) {};
		\node [style=none] (37) at (4.637, 0.3) {};
		\node [style=none] (38) at (4.637, 0.9) {};
		\node [style=none] (39) at (0.875, 0) {$\frac{e^{-i\pi/4}}{\sqrt2}$};
	\end{pgfonlayer}
	\begin{pgfonlayer}{edgelayer}
		\draw (4.center) to (3);
		\draw [style=hadamard edge, in=165, out=-75] (6) to (3);
		\draw (8.center) to (7);
		\draw (11.center) to (10);
		\draw (14.center) to (13);
		\draw [style=hadamard edge, in=180, out=-45] (6) to (7);
		\draw [style=hadamard edge, in=45, out=180] (10) to (6);
		\draw [style=hadamard edge, in=-165, out=75] (6) to (13);
		\draw [in=165, out=-75] (22) to (21.center);
		\draw [in=180, out=-45] (22) to (23.center);
		\draw [in=45, out=180] (24.center) to (22);
		\draw [in=-165, out=75] (22) to (25.center);
		\draw [style=hadamard edge, in=165, out=-75] (31) to (30);
		\draw [style=hadamard edge, in=180, out=-45] (31) to (32);
		\draw [style=hadamard edge, in=45, out=180] (33) to (31);
		\draw [style=hadamard edge, in=-165, out=75] (31) to (34);
		\draw (35.center) to (30);
		\draw (36.center) to (32);
		\draw (37.center) to (33);
		\draw (38.center) to (34);
	\end{pgfonlayer}
\end{tikzpicture}

%% file: figures/cat-3-decomp.tikz
\begin{tikzpicture}[scale=2]
	\begin{pgfonlayer}{nodelayer}
		\node [style=Z dot] (3) at (-1.425, 0) {};
		\node [style=Z phase dot] (4) at (-0.788, 0.625) {$\frac\pi4$};
		\node [style=none] (5) at (-0.437, 0.625) {};
		\node [style=Z phase dot] (7) at (-0.788, 0) {$\frac\pi4$};
		\node [style=none] (8) at (-0.437, 0) {};
		\node [style=Z phase dot] (10) at (-0.788, -0.625) {$\frac\pi4$};
		\node [style=none] (11) at (-0.437, -0.625) {};
		\node [style=none] (13) at (0, 0) {$=$};
		\node [style=Z dot] (14) at (2.112, 0.75) {};
		\node [style=Z phase dot] (15) at (1.725, 0.075) {$~-\frac\pi2~$};
		\node [style=none] (16) at (2.537, 0.375) {};
		\node [style=none] (17) at (2.537, -0.25) {};
		\node [style=none] (18) at (2.537, -0.875) {};
		\node [style=none] (20) at (3.125, 0) {$+~i$};
		\node [style=Z dot] (21) at (4.062, 0.75) {};
		\node [style=Z dot] (22) at (3.675, 0.075) {};
		\node [style=Z dot] (23) at (4.312, 0.375) {};
		\node [style=Z dot] (24) at (4.312, -0.25) {};
		\node [style=Z dot] (25) at (4.312, -0.875) {};
		\node [style=none] (26) at (4.55, 0.375) {};
		\node [style=none] (27) at (4.55, -0.25) {};
		\node [style=none] (28) at (4.55, -0.875) {};
		\node [style=none] (29) at (0.825, 0) {$\frac{e^{-i\pi/4}}{\sqrt2}$};
	\end{pgfonlayer}
	\begin{pgfonlayer}{edgelayer}
		\draw (5.center) to (4);
		\draw (8.center) to (7);
		\draw (11.center) to (10);
		\draw [style=hadamard edge, in=-165, out=75] (3) to (4);
		\draw [style=hadamard edge] (7) to (3);
		\draw [style=hadamard edge, in=165, out=-75] (3) to (10);
		\draw [style=hadamard edge, in=-165, out=90] (15) to (14);
		\draw [in=-180, out=45] (15) to (16.center);
		\draw [in=-45, out=-180] (17.center) to (15);
		\draw [in=165, out=-75] (15) to (18.center);
		\draw [in=-165, out=90] (22) to (21);
		\draw [style=hadamard edge, in=-180, out=45] (22) to (23);
		\draw [style=hadamard edge, in=-45, out=-180] (24) to (22);
		\draw [style=hadamard edge, in=165, out=-75] (22) to (25);
		\draw (23) to (26.center);
		\draw (24) to (27.center);
		\draw (25) to (28.center);
	\end{pgfonlayer}
\end{tikzpicture}

%% file: figures/cat-5-decomp.tikz
\begin{tikzpicture}[xscale=1.7, yscale=2]
	\begin{pgfonlayer}{nodelayer}
		\node [style=Z phase dot] (58) at (-1.113, -1.425) {$\frac\pi4$};
		\node [style=none] (59) at (-0.512, -1.425) {};
		\node [style=Z phase dot] (60) at (-1.113, -0.825) {$\frac\pi4$};
		\node [style=none] (61) at (-0.512, -0.825) {};
		\node [style=Z dot] (62) at (-1.75, 0.05) {};
		\node [style=Z phase dot] (63) at (-1.113, -0.225) {$\frac\pi4$};
		\node [style=none] (64) at (-0.512, -0.225) {};
		\node [style=Z phase dot] (65) at (-1.113, 0.35) {$\frac\pi4$};
		\node [style=none] (66) at (-0.512, 0.35) {};
		\node [style=Z phase dot] (67) at (-1.113, 0.95) {$\frac\pi4$};
		\node [style=none] (68) at (-0.512, 0.95) {};
		\node [style=none] (71) at (0, 0) {$=$};
		\node [style=none] (72) at (2.087, -1.45) {};
		\node [style=none] (73) at (2.087, -0.85) {};
		\node [style=Z phase dot] (74) at (1.3, 0.025) {$~-\frac\pi2~$};
		\node [style=none] (75) at (2.087, -0.25) {};
		\node [style=none] (76) at (2.087, 0.325) {};
		\node [style=none] (77) at (2.087, 0.925) {};
		\node [style=Z dot] (78) at (2.087, 1.475) {};
		\node [style=none] (79) at (3.25, 0) {$+~\frac{i e^{i\pi/4}}{\sqrt2}$};
		\node [style=Z dot] (80) at (4.987, -1.425) {};
		\node [style=Z dot] (81) at (4.987, -0.825) {};
		\node [style=Z dot] (82) at (4.35, 0.05) {};
		\node [style=Z dot] (83) at (4.987, -0.225) {};
		\node [style=Z dot] (84) at (4.987, 0.35) {};
		\node [style=Z dot] (85) at (4.987, 0.95) {};
		\node [style=Z dot] (86) at (4.987, 1.5) {};
		\node [style=none] (87) at (6.225, 0) {$-\ \frac{e^{i\pi/4}}{\sqrt{2}}$};
		\node [style=none] (88) at (5.362, -1.425) {};
		\node [style=none] (89) at (5.362, -0.825) {};
		\node [style=none] (90) at (5.362, -0.225) {};
		\node [style=none] (91) at (5.362, 0.35) {};
		\node [style=none] (92) at (5.362, 0.95) {};
		\node [style=none] (94) at (0.5, 0) {$\frac12$};
		\node [style=Z phase dot] (95) at (7.987, -1.425) {$\frac\pi2$};
		\node [style=Z phase dot] (96) at (7.987, -0.825) {$\frac\pi2$};
		\node [style=Z dot] (97) at (7.1, 0.05) {};
		\node [style=Z phase dot] (98) at (7.987, -0.225) {$\frac\pi2$};
		\node [style=Z phase dot] (99) at (7.987, 0.35) {$\frac\pi2$};
		\node [style=Z phase dot] (100) at (7.987, 0.95) {$\frac\pi2$};
		\node [style=Z dot] (101) at (7.987, 1.5) {};
		\node [style=none] (102) at (8.487, -1.425) {};
		\node [style=none] (103) at (8.487, -0.825) {};
		\node [style=none] (104) at (8.487, -0.225) {};
		\node [style=none] (105) at (8.487, 0.35) {};
		\node [style=none] (106) at (8.487, 0.95) {};
	\end{pgfonlayer}
	\begin{pgfonlayer}{edgelayer}
		\draw (59.center) to (58);
		\draw (61.center) to (60);
		\draw [style=hadamard edge, in=165, out=-90] (62) to (58);
		\draw [style=hadamard edge, in=165, out=-75] (62) to (60);
		\draw (64.center) to (63);
		\draw (66.center) to (65);
		\draw (68.center) to (67);
		\draw [style=hadamard edge, in=180, out=-45] (62) to (63);
		\draw [style=hadamard edge, in=45, out=180] (65) to (62);
		\draw [style=hadamard edge, in=-165, out=75] (62) to (67);
		\draw [in=165, out=-90] (74) to (72.center);
		\draw [in=165, out=-75] (74) to (73.center);
		\draw [in=180, out=-45] (74) to (75.center);
		\draw [in=45, out=180] (76.center) to (74);
		\draw [in=-165, out=75] (74) to (77.center);
		\draw [style=hadamard edge, in=-165, out=90] (74) to (78);
		\draw [style=hadamard edge, in=165, out=-90] (82) to (80);
		\draw [style=hadamard edge, in=165, out=-75] (82) to (81);
		\draw [style=hadamard edge, in=180, out=-45] (82) to (83);
		\draw [style=hadamard edge, in=45, out=180] (84) to (82);
		\draw [style=hadamard edge, in=-165, out=75] (82) to (85);
		\draw [in=-165, out=90] (82) to (86);
		\draw (80) to (88.center);
		\draw (81) to (89.center);
		\draw (83) to (90.center);
		\draw (84) to (91.center);
		\draw (85) to (92.center);
		\draw [style=hadamard edge, in=165, out=-90] (97) to (95);
		\draw [style=hadamard edge, in=165, out=-75] (97) to (96);
		\draw [style=hadamard edge, in=180, out=-45] (97) to (98);
		\draw [style=hadamard edge, in=45, out=180] (99) to (97);
		\draw [style=hadamard edge, in=-165, out=75] (97) to (100);
		\draw [in=-165, out=90] (97) to (101);
		\draw (95) to (102.center);
		\draw (96) to (103.center);
		\draw (98) to (104.center);
		\draw (99) to (105.center);
		\draw (100) to (106.center);
	\end{pgfonlayer}
\end{tikzpicture}

%% file: figures/pi-cat-state.tikz
\begin{tikzpicture}[scale=2]
	\begin{pgfonlayer}{nodelayer}
		\node [style=Z phase dot] (0) at (-1.613, -0.925) {$\frac\pi4$};
		\node [style=none] (1) at (-1.012, -0.925) {};
		\node [style=Z phase dot] (3) at (-1.613, -0.25) {$\frac\pi4$};
		\node [style=none] (4) at (-1.012, -0.25) {};
		\node [style=Z phase dot] (6) at (-1.613, 0.725) {$\frac\pi4$};
		\node [style=none] (7) at (-1.012, 0.725) {};
		\node [style=Z phase dot] (9) at (-2.25, 0) {$\pi$};
		\node [style=none] (10) at (-1.6, 0.325) {$\vdots$};
		\node [style=none] (12) at (-0.5, 0) {$=$};
		\node [style=Z phase dot] (13) at (1.387, -0.925) {$\frac\pi4$};
		\node [style=none] (14) at (3.538, -0.925) {};
		\node [style=Z phase dot] (16) at (1.387, -0.25) {$\frac\pi4$};
		\node [style=none] (17) at (2.788, -0.25) {};
		\node [style=Z phase dot] (19) at (1.387, 0.725) {$\frac\pi4$};
		\node [style=none] (20) at (2.788, 0.725) {};
		\node [style=Z dot] (22) at (0.75, 0) {};
		\node [style=none] (23) at (1.4, 0.325) {$\vdots$};
		\node [style=Z phase dot] (25) at (2.75, -0.925) {$\pi$};
		\node [style=none] (27) at (0.25, 0) {$e^{i\frac\pi4}$};
		\node [style=Z dot] (28) at (3.25, -0.925) {};
		\node [style=Z phase dot] (29) at (2.038, -0.925) {$~-\frac\pi2~$};
	\end{pgfonlayer}
	\begin{pgfonlayer}{edgelayer}
		\draw (1.center) to (0);
		\draw (4.center) to (3);
		\draw (7.center) to (6);
		\draw [style=hadamard edge, in=165, out=-90] (9) to (0);
		\draw [style=hadamard edge, in=180, out=-45] (9) to (3);
		\draw [style=hadamard edge, in=-165, out=90] (9) to (6);
		\draw (17.center) to (16);
		\draw (20.center) to (19);
		\draw [style=hadamard edge, in=165, out=-90] (22) to (13);
		\draw [style=hadamard edge, in=180, out=-45] (22) to (16);
		\draw [style=hadamard edge, in=-165, out=90] (22) to (19);
		\draw (28) to (14.center);
		\draw [style=hadamard edge] (28) to (25);
		\draw [style=hadamard edge] (29) to (25);
		\draw (13) to (29);
	\end{pgfonlayer}
\end{tikzpicture}

%% file: figures/phase-gadget-to-cat.tikz
\begin{tikzpicture}[scale=2]
	\begin{pgfonlayer}{nodelayer}
		\node [style=Z dot] (0) at (-3, 0.25) {};
		\node [style=Z phase dot] (1) at (-3, 0.75) {$~(2k{+}1)\frac\pi4~$};
		\node [style=Z phase dot] (2) at (-4, -0.25) {$~(2k_1{+}1)\frac\pi4~$};
		\node [style=Z phase dot] (3) at (-2, -0.25) {$~(2k_n{+}1)\frac\pi4~$};
		\node [style=none] (4) at (-5, 0) {};
		\node [style=none] (5) at (-5, -0.5) {};
		\node [style=none] (6) at (-3, -0.25) {...};
		\node [style=none] (7) at (-1, -0.5) {};
		\node [style=none] (8) at (-1, 0) {};
		\node [style=none] (9) at (-1, -0.15) {$\vdots$};
		\node [style=none] (10) at (-5, -0.15) {$\vdots$};
		\node [style=none] (11) at (-0.25, 0) {$=$};
		\node [style=Z dot] (12) at (2.5, 1.25) {};
		\node [style=Z phase dot] (14) at (1.5, -0.5) {$~k_1\frac\pi2~$};
		\node [style=Z phase dot] (15) at (3.5, -0.5) {$~k_n\frac\pi2~$};
		\node [style=none] (16) at (0.5, -0.25) {};
		\node [style=none] (17) at (0.5, -0.75) {};
		\node [style=none] (18) at (2.5, -0.5) {...};
		\node [style=none] (19) at (4.5, -0.75) {};
		\node [style=none] (20) at (4.5, -0.25) {};
		\node [style=none] (21) at (4.5, -0.4) {$\vdots$};
		\node [style=none] (22) at (0.5, -0.4) {$\vdots$};
		\node [style=Z phase dot] (23) at (1.5, 0.5) {$\frac\pi4$};
		\node [style=Z phase dot] (24) at (3.5, 0.5) {$\frac\pi4$};
		\node [style=Z phase dot] (25) at (1, 0.5) {$\frac\pi4$};
		\node [style=Z phase dot] (26) at (1, 0) {$~k\frac\pi2~$};
		\node [style=none] (27) at (2.5, 0.5) {...};
	\end{pgfonlayer}
	\begin{pgfonlayer}{edgelayer}
		\draw [style=hadamard edge] (1) to (0);
		\draw [style=hadamard edge] (0) to (2);
		\draw (2) to (4.center);
		\draw (2) to (5.center);
		\draw [style=hadamard edge] (0) to (3);
		\draw (8.center) to (3);
		\draw (3) to (7.center);
		\draw (14) to (16.center);
		\draw (14) to (17.center);
		\draw (20.center) to (15);
		\draw (15) to (19.center);
		\draw (23) to (14);
		\draw (24) to (15);
		\draw (25) to (26);
		\draw [style=hadamard edge, in=-180, out=75] (25) to (12);
		\draw [style=hadamard edge, in=75, out=-165] (12) to (23);
		\draw [style=hadamard edge, in=105, out=0] (12) to (24);
	\end{pgfonlayer}
\end{tikzpicture}

%% file: figures/cat-n.tikz
\begin{tikzpicture}[scale=2]
	\begin{pgfonlayer}{nodelayer}
		\node [style=none] (0) at (-8.5, 0.375) {$\text{cat}_{4k+2}$};
		\node [style=none] (1) at (-9.5, 0) {};
		\node [style=none] (2) at (-7.5, 0) {};
		\node [style=none] (3) at (-8.5, 1.25) {};
		\node [style=none] (4) at (-8.5, 0) {};
		\node [style=none] (5) at (-8.5, -1) {};
		\node [style=none] (6) at (-8.625, -0.55) {};
		\node [style=none] (7) at (-8.375, -0.45) {};
		\node [style=none, font={\scriptsize}, anchor=west] (8) at (-8.25, -0.5) {$4k{+}2$};
		\node [style=none] (9) at (-7, 0) {$=$};
		\node [style=none] (10) at (-5.75, 0.375) {$\text{cat}_6$};
		\node [style=none] (11) at (-6.5, 0) {};
		\node [style=none] (12) at (-5, 0) {};
		\node [style=none] (13) at (-5.75, 1) {};
		\node [style=none] (14) at (-6, 0) {};
		\node [style=none] (15) at (-6, -1) {};
		\node [style=none] (16) at (-6.125, -0.55) {};
		\node [style=none] (17) at (-5.875, -0.45) {};
		\node [style=none, font={\scriptsize}, anchor=west] (18) at (-5.75, -0.5) {$5$};
		\node [style=none] (19) at (-4, 0.375) {$\text{cat}_6$};
		\node [style=none] (20) at (-4.75, 0) {};
		\node [style=none] (21) at (-3.25, 0) {};
		\node [style=none] (22) at (-4, 1) {};
		\node [style=none] (23) at (-4, 0) {};
		\node [style=none] (24) at (-4, -1) {};
		\node [style=none] (25) at (-4.125, -0.55) {};
		\node [style=none] (26) at (-3.875, -0.45) {};
		\node [style=none, font={\scriptsize}, anchor=west] (27) at (-3.75, -0.5) {$4$};
		\node [style=none] (28) at (-1.525, 0.375) {$\text{cat}_6$};
		\node [style=none] (29) at (-2.275, 0) {};
		\node [style=none] (30) at (-0.775, 0) {};
		\node [style=none] (31) at (-1.525, 1) {};
		\node [style=none] (32) at (-1.275, 0) {};
		\node [style=none] (33) at (-1.275, -1) {};
		\node [style=none] (34) at (-1.4, -0.55) {};
		\node [style=none] (35) at (-1.15, -0.45) {};
		\node [style=none, font={\scriptsize}, anchor=west] (36) at (-1.025, -0.5) {$5$};
		\node [style=none] (37) at (-5.5, 0) {};
		\node [style=none] (38) at (-4.375, 0) {};
		\node [style=none] (39) at (-3.625, 0) {};
		\node [style=none] (40) at (-3.5, -0.25) {};
		\node [style=none] (41) at (-1.775, 0) {};
		\node [style=none] (43) at (-2.95, 0) {...};
		\node [style=none] (44) at (-4.925, -0.45) {$\text{cat}_2$};
		\node [style=none] (45) at (-5.55, -0.25) {};
		\node [style=none] (46) at (-4.3, -0.25) {};
		\node [style=none] (47) at (-4.925, -1) {};
		\node [style=none] (48) at (-5.175, -0.25) {};
		\node [style=none] (49) at (-4.675, -0.25) {};
		\node [style=none] (50) at (-2.15, -0.45) {$\text{cat}_2$};
		\node [style=none] (51) at (-2.775, -0.25) {};
		\node [style=none] (52) at (-1.525, -0.25) {};
		\node [style=none] (53) at (-2.15, -1) {};
		\node [style=none] (54) at (-2.4, -0.25) {};
		\node [style=none] (55) at (-1.9, -0.25) {};
		\node [style=none] (56) at (-2.5, 0) {};
	\end{pgfonlayer}
	\begin{pgfonlayer}{edgelayer}
		\draw (3.center) to (1.center);
		\draw (1.center) to (2.center);
		\draw (2.center) to (3.center);
		\draw (4.center) to (5.center);
		\draw (7.center) to (6.center);
		\draw (13.center) to (11.center);
		\draw (11.center) to (12.center);
		\draw (12.center) to (13.center);
		\draw (14.center) to (15.center);
		\draw (17.center) to (16.center);
		\draw (22.center) to (20.center);
		\draw (20.center) to (21.center);
		\draw (21.center) to (22.center);
		\draw (23.center) to (24.center);
		\draw (26.center) to (25.center);
		\draw (31.center) to (29.center);
		\draw (29.center) to (30.center);
		\draw (30.center) to (31.center);
		\draw (32.center) to (33.center);
		\draw (35.center) to (34.center);
		\draw (40.center) to (39.center);
		\draw (47.center) to (45.center);
		\draw (45.center) to (46.center);
		\draw (46.center) to (47.center);
		\draw (48.center) to (37.center);
		\draw (49.center) to (38.center);
		\draw (53.center) to (51.center);
		\draw (51.center) to (52.center);
		\draw (52.center) to (53.center);
		\draw (55.center) to (41.center);
		\draw (56.center) to (54.center);
	\end{pgfonlayer}
\end{tikzpicture}

%% file: figures/magic-state-5-decomp.tikz
\begin{tikzpicture}[scale=2]
	\begin{pgfonlayer}{nodelayer}
		\node [style=Z phase dot] (0) at (1.812, 1.55) {$\frac\pi4$};
		\node [style=Z phase dot] (1) at (2.613, 1.55) {$~-\frac\pi4~$};
		\node [style=Z phase dot] (3) at (1.812, 0.925) {$\frac\pi4$};
		\node [style=none] (4) at (2.163, 0.925) {};
		\node [style=Z dot] (6) at (1.175, 0) {};
		\node [style=Z phase dot] (7) at (1.812, 0.3) {$\frac\pi4$};
		\node [style=none] (8) at (2.163, 0.3) {};
		\node [style=Z phase dot] (10) at (1.812, -0.325) {$\frac\pi4$};
		\node [style=none] (11) at (2.163, -0.325) {};
		\node [style=Z phase dot] (13) at (1.812, -0.95) {$\frac\pi4$};
		\node [style=none] (14) at (2.163, -0.95) {};
		\node [style=Z phase dot] (16) at (1.812, -1.575) {$\frac\pi4$};
		\node [style=none] (17) at (2.163, -1.575) {};
		\node [style=none] (98) at (0.05, 0) {$=$};
		\node [style=Z phase dot] (105) at (-0.863, -1.625) {$\frac\pi4$};
		\node [style=none] (106) at (-0.337, -1.625) {};
		\node [style=Z phase dot] (107) at (-0.863, -0.875) {$\frac\pi4$};
		\node [style=none] (108) at (-0.337, -0.875) {};
		\node [style=Z phase dot] (109) at (-0.863, -0.125) {$\frac\pi4$};
		\node [style=none] (110) at (-0.337, -0.125) {};
		\node [style=Z phase dot] (111) at (-0.863, 0.625) {$\frac\pi4$};
		\node [style=none] (112) at (-0.337, 0.625) {};
		\node [style=Z phase dot] (113) at (-0.863, 1.375) {$\frac\pi4$};
		\node [style=none] (114) at (-0.337, 1.375) {};
		\node [style=none] (120) at (0.6, 0) {$4$};
		\node [style=none] (134) at (5.062, -1.525) {};
		\node [style=none] (135) at (5.062, -0.9) {};
		\node [style=Z phase dot] (136) at (4.175, 0.025) {$~-\frac\pi2~$};
		\node [style=none] (137) at (5.062, -0.275) {};
		\node [style=none] (138) at (5.062, 0.35) {};
		\node [style=none] (139) at (5.062, 0.975) {};
		\node [style=Z phase dot] (140) at (5.062, 1.6) {$\ -\frac\pi4~$};
		\node [style=none] (141) at (6.3, 0) {$+~2\sqrt{2}i e^{i\pi/4}$};
		\node [style=Z dot] (142) at (8.312, -1.5) {};
		\node [style=Z dot] (143) at (8.312, -0.875) {};
		\node [style=Z dot] (144) at (7.675, 0.05) {};
		\node [style=Z dot] (145) at (8.312, -0.25) {};
		\node [style=Z dot] (146) at (8.312, 0.375) {};
		\node [style=Z dot] (147) at (8.312, 1) {};
		\node [style=Z phase dot] (148) at (8.312, 1.625) {$\ -\frac\pi4~$};
		\node [style=none] (149) at (9.975, 0) {$-\ 2\sqrt{2} e^{i\pi/4}$};
		\node [style=none] (150) at (8.687, -1.5) {};
		\node [style=none] (151) at (8.687, -0.875) {};
		\node [style=none] (152) at (8.687, -0.25) {};
		\node [style=none] (153) at (8.687, 0.375) {};
		\node [style=none] (154) at (8.687, 1) {};
		\node [style=none] (155) at (3.55, 0) {$2$};
		\node [style=Z phase dot] (156) at (11.987, -1.5) {$\frac\pi2$};
		\node [style=Z phase dot] (157) at (11.987, -0.875) {$\frac\pi2$};
		\node [style=Z dot] (158) at (11.1, 0.05) {};
		\node [style=Z phase dot] (159) at (11.987, -0.25) {$\frac\pi2$};
		\node [style=Z phase dot] (160) at (11.987, 0.375) {$\frac\pi2$};
		\node [style=Z phase dot] (161) at (11.987, 1) {$\frac\pi2$};
		\node [style=Z phase dot] (162) at (11.987, 1.625) {$\frac\pi4$};
		\node [style=none] (163) at (12.487, -1.5) {};
		\node [style=none] (164) at (12.487, -0.875) {};
		\node [style=none] (165) at (12.487, -0.25) {};
		\node [style=none] (166) at (12.487, 0.375) {};
		\node [style=none] (167) at (12.487, 1) {};
		\node [style=none] (168) at (2.975, 0) {=};
	\end{pgfonlayer}
	\begin{pgfonlayer}{edgelayer}
		\draw (1) to (0);
		\draw (4.center) to (3);
		\draw [style=hadamard edge, in=-165, out=90] (6) to (0);
		\draw [style=hadamard edge, in=-165, out=75] (6) to (3);
		\draw (8.center) to (7);
		\draw (11.center) to (10);
		\draw (14.center) to (13);
		\draw (17.center) to (16);
		\draw [style=hadamard edge, in=-180, out=45] (6) to (7);
		\draw [style=hadamard edge, in=-45, out=-180] (10) to (6);
		\draw [style=hadamard edge, in=165, out=-75] (6) to (13);
		\draw [style=hadamard edge, in=165, out=-90] (6) to (16);
		\draw (106.center) to (105);
		\draw (108.center) to (107);
		\draw (110.center) to (109);
		\draw (112.center) to (111);
		\draw (114.center) to (113);
		\draw [in=165, out=-90] (136) to (134.center);
		\draw [in=165, out=-75] (136) to (135.center);
		\draw [in=180, out=-45] (136) to (137.center);
		\draw [in=45, out=180] (138.center) to (136);
		\draw [in=-165, out=75] (136) to (139.center);
		\draw [in=-165, out=90] (136) to (140);
		\draw [style=hadamard edge, in=165, out=-90] (144) to (142);
		\draw [style=hadamard edge, in=165, out=-75] (144) to (143);
		\draw [style=hadamard edge, in=180, out=-45] (144) to (145);
		\draw [style=hadamard edge, in=45, out=180] (146) to (144);
		\draw [style=hadamard edge, in=-165, out=75] (144) to (147);
		\draw [style=hadamard edge, in=-165, out=90] (144) to (148);
		\draw (142) to (150.center);
		\draw (143) to (151.center);
		\draw (145) to (152.center);
		\draw (146) to (153.center);
		\draw (147) to (154.center);
		\draw [style=hadamard edge, in=165, out=-90] (158) to (156);
		\draw [style=hadamard edge, in=165, out=-75] (158) to (157);
		\draw [style=hadamard edge, in=180, out=-45] (158) to (159);
		\draw [style=hadamard edge, in=45, out=180] (160) to (158);
		\draw [style=hadamard edge, in=-165, out=75] (158) to (161);
		\draw [style=hadamard edge, in=-165, out=90] (158) to (162);
		\draw (156) to (163.center);
		\draw (157) to (164.center);
		\draw (159) to (165.center);
		\draw (160) to (166.center);
		\draw (161) to (167.center);
	\end{pgfonlayer}
\end{tikzpicture}